\documentclass[pre,twocolumn,english,aps,a4paper,floatfix]{revtex4}
\usepackage{color}
\usepackage[latin1]{inputenc}
\usepackage{graphicx}
\usepackage{bm}
\usepackage{epsfig}
\usepackage{amsmath}
\usepackage{amssymb}
\usepackage[svgnames]{xcolor}
\usepackage{todonotes}
\usepackage[hidelinks]{hyperref}
\usepackage{float}

\begin{document}

\title{Biaxial nematics and nematic-nematic demixing in polydisperse mixtures of hard board-like particle fluids}
\author{Yuri Mart\'{\i}nez-Rat\'on}  
\email{yuri@math.uc3m.es}
\affiliation{Universidad Carlos III de Madrid, Departamento de Matem\'aticas, Grupo Interdisciplinar 
de Sistemas Complejos (GISC), 28911 Legan\'es, Spain.}
\author{Daniel de las Heras}
\affiliation{Institut f\"ur Theoretische Physik, Universit\"at T\"ubingen, D-72076 T\"ubingen, Germany}
\author{Enrique Velasco}
\affiliation{Departamento de F\'{\i}sica Te\'orica de la Materia Condensada, Instituto de 
F\'{\i}sica de la Materia Condensada (IFIMAC) and Instituto 
de Ciencias de Materiales Nicol\'as Cabrera, Universidad Aut\'onoma de Madrid, E-28049 Madrid, Spain}

\begin{abstract}
We study the bulk phase behavior of a polydisperse liquid crystal fluid made of biaxial boards with restricted orientations using both fundamental measure theory and Monte Carlo computer simulations.
The continuous polydispersity is included in the intermediate particle length of the boards via a truncated Schulz distribution.
By calculating several phase diagrams across a range of polydispersity coefficients, we find that polydispersity (i) enhances demixing between two uniaxial nematic phases and (ii) expands the stability region of the biaxial phase.
Although metastable with respect to non-uniform phases, we identify certain mixtures exhibiting two-phase coexistence paths involving uniaxial-uniaxial and uniaxial-biaxial phase separations. Whether these paths can be completed depends on the precise shape of the parent distribution function.
Monte Carlo simulations performed for a representative case exhibit the same phase diagram topology predicted by theory, thereby validating the theoretical approach.
The main difference between both approaches lies in quantitative agreement, with simulated phase transitions systematically occurring at higher packing fractions than those predicted theoretically.
Our combined theoretical and simulation results may prove relevant to the design and interpretation of sedimentation experiments on colloidal suspensions of polydisperse anisotropic particles.
\end{abstract}

\maketitle

\section{Introduction}

In a biaxial nematic (B) liquid-crystal phase, the particles exhibit orientational order along two perpendicular axes while maintaining complete positional disorder.
Since the original prediction of the B-phase by Freiser \cite{Freiser}, and the first theoretical models demonstrating its possible stability \cite{Alben1}, the first experimental realization was achieved in a lyotropic ternary liquid-crystal mixture \cite{Yu-Saupe1,Yu-Saupe2}. Despite the large body of theoretical and simulation studies devoted to thermotropic and colloidal liquid crystals with either attractive-repulsive or purely hard-core interactions \cite{Taylor,Berardi1,Bates1,M-R1,Vanakaras2,Roij1}, unambiguous experimental evidence for the B-phase in these systems has emerged only relatively recently, both in molecular systems \cite{Kumar,Severing,Madsen} and in colloidal suspensions \cite{van_den_Pol}.

Much of the theoretical work has focused on the role of particle geometry in stabilizing the B-phase, considering systems of boomerang- or V-shaped particles \cite{Bates1} as well as board-like particles \cite{Vanakaras2,M-R1,Roij1}. More recently, other geometries, including triangular and rhomboidal prisms, have also been predicted to exhibit stable biaxial nematic phases \cite{Roij1}. The considerable interest in this topic is reflected in the large number of theoretical, simulation, and experimental studies, as well as in several review articles (see Refs. \cite{Review1,Review2,Review3,Review4} for an extensive list of relevant papers).

The most extensively studied particle geometry consists of hard board-like particles with edge lengths satisfying
$\sigma_1\geq \sigma_2\geq \sigma_3$. Early theoretical studies focused on particles with dimensions close to the so-called \textit{dual shape}, defined by $\sigma_2=\sqrt{\sigma_1\sigma_3}$, but considering only moderate values of the largest aspect ratio
$\kappa_1=\sigma_1/\sigma_3$. Under these conditions, the B-phase was found not to be stable \cite{Cuetos1}. It was only when substantially larger aspect ratios $\kappa_1$ were explored that a stable B-phase was predicted for $\kappa_1\gtrsim 23$ \cite{Roij1}.

Another important avenue of research, inspired by the pioneering work of Alben \cite{Alben2}, focused on mixtures of hard rods and plates. Theoretical studies predicted the stability of the B-phase over specific ranges of composition, density, and particle aspect ratios \cite{Bolhuis,Wensink1,Varga1,Photinos1,Cuetos2,M-R2,M-R3}. However, despite several experimental investigations of colloidal rod-plate mixtures with quasi-hard-core interactions, no biaxial nematic phase has been observed in these systems \cite{Kooij1,Kooij2}.

Particle size-polydispersity has been identified as an important factor affecting the stability of the B-phase, primarily because it tends to destabilize spatially ordered phases in favour of uniform liquid-crystalline phases \cite{M-R2,Patti1,Patti2}. Nevertheless, theoretical studies indicate that enhancing the stability of the biaxial nematic phase requires not only an appropriate degree of polydispersity, but also a carefully tailored size distribution, in order to avoid a competing nematic demixing. Such precise control over the particle-size distribution is difficult to achieve experimentally.

In addition to its influence on the stability of the biaxial nematic phase, size polydispersity has profound effects on the phase behaviour of liquid-crystalline systems. For example, in fluids of length-polydisperse hard rods, the smectic phase becomes unstable above a terminal degree of polydispersity \cite{Bates2}, whereas in suspensions of hard platelets, size polydispersity leads to strong fractionation between coexisting phases \cite{Bates3}. In rod-plate mixtures, polydispersity in particle shape has also been shown to stabilize the biaxial nematic phase, while the corresponding monodisperse binary mixtures with the same average particle aspect ratios do not exhibit a stable B-phase \cite{M-R2}.

The strong fractionation of particle sizes between coexisting liquid-crystalline phases, together with the selective destabilization of spatially ordered phases, can substantially modify bulk phase equilibria \cite{Lekkerkerker,Sollich1,Sollich2,Petukhov}. In some cases, it even reverses the densities of the coexisting isotropic (I) and nematic (N) phases \cite{Wensink2}. Conversely, size polydispersity may induce nematic-nematic (N-N) demixing, an effect commonly observed in binary mixtures that significantly reduces the stability region of the B-phase \cite{Wensink1,Varga1,Bolhuis,Wensink3,Bela,M-R3}.

Finally, both theoretical \cite{Eckert1,Eckert2,Eckert3} and experimental \cite{Kooij1,Kooij2,Lekkerkerker} studies have revealed an extraordinary variety of phase-stacking sequences in sedimented columns of polydisperse colloidal liquid crystals. This rich behaviour arises from the interplay between size polydispersity, the external gravitational field, and the intrinsic complexity of the bulk phase diagram.

Incorporating continuous size polydispersity into theoretical models of liquid crystals is a challenging task because it requires describing the orientational and spatial ordering of an infinite number of particle species. Consequently, the number of degrees of freedom becomes formally infinite, making the exact treatment of the problem intractable. Several approximation schemes have therefore been developed to reduce its complexity.

A first approach consists of replacing the continuous size distribution by a finite, albeit possibly large, number of discrete species. A second approximation reduces the orientational degrees of freedom, for example through the Zwanzig model \cite{Zwanzig1,Zwanzig2}, in which particle orientations are restricted to the three Cartesian axes. A third and particularly powerful approach is the moment method, whereby the infinite-dimensional distribution function is projected onto a finite set of generalized moments \cite{Sollich3,Sollich4}. 
This is particularly appealing because it retains a continuous description of polydispersity while reducing the thermodynamic problem to a finite number of variables.
If the excess free-energy functional depends only on these generalized moments, the phase behaviour can be determined using the standard tools of equilibrium thermodynamics to calculate phase coexistence, cloud and shadow curves, spinodal instabilities, and related properties. The moment method is exact for cloud-shadow coexistence and spinodal calculations, whereas for general multiphase coexistence it provides an accurate approximation.

These approximations may have a significant impact on the theoretical predictions. For example, it is well known that the Zwanzig approximation overestimates the stability of the biaxial nematic phase in systems of hard board-like particles, predicting its appearance at substantially lower aspect ratios than those obtained from freely rotating models \cite{M-R3,Bela}. Nevertheless, the Zwanzig model has proved remarkably successful in providing qualitative predictions of liquid-crystal phase behaviour \cite{M-R5}. Monte Carlo (MC) simulations of Zwanzig models are comparatively scarce. Moreover, most studies have been performed on lattice models and restricted to uniaxial board-like particles ($\sigma_1=\sigma_2$) \cite{Harrowell,Oettel2,Ramirez-Pastor}.

In this work, we investigate the effect of continuous polydispersity in the intermediate edge length, $\sigma_2$, of hard board-like particles on the topology of the phase diagram. Particular attention is devoted to the influence of polydispersity on the nematic-nematic (N-N) demixing region and on the stability of the biaxial nematic phase.

To this end, we consider two values of the largest particle dimension, $\sigma_1=\{5,10\}$, while fixing the smallest edge length to $\sigma_3=1$, which sets the length scale of the model. The continuous size distribution is described by a truncated unimodal Schulz distribution, and several values of the polydispersity coefficient are investigated.

We extend the Fundamental Measure Theory (FMT) density functional developed for monodisperse hard board-like particles in the Zwanzig approximation \cite{M-R1} to continuously polydisperse systems and implement it numerically to determine the corresponding phase diagrams. To assess the reliability of the theoretical predictions, we also perform MC simulations in the canonical (NVT) ensemble for selected cases.

We find that increasing polydispersity widens the demixing gap between the rod-rich (N$_{\rm r}$) and plate-rich (N$_{\rm p}$) nematic phases. This effect, however, is less pronounced than the corresponding expansion of the stability region of the biaxial nematic phase, which is located above the N$_{\rm r}$-N$_{\rm p}$ demixing gap and below the stability region of the smectic (Sm) phase. The position of the multicritical point, where the N$_{\rm r}$-N$_{\rm p}$ demixing line meets the isotropic-nematic I-N$_{{\rm r,p}}$ coexistence, is strongly affected by polydispersity, reflecting the dominant influence of plate-like particles on the phase behaviour.

We also find that the Sm phases of both rods and plates can accommodate relatively high degrees of polydispersity. This is because the density modulation always develops along a direction parallel to one of the monodisperse particle dimensions, namely $\sigma_1$ (for rods) or $\sigma_3$ (for plates), while the polydisperse dimension, $\sigma_2$, lies within the smectic layers.

MC simulations performed for particles with $\sigma_1=10$ and the highest polydispersity considered confirm the overall topology of the theoretically predicted phase diagram.

For $\sigma_1=5$, the theoretical calculations indicate that the B-phase remains unstable even at the largest polydispersities investigated. In this case, only a relatively narrow N$_{\rm r}$-N$_{\rm p}$ demixing region is found above the multicritical point and below the stability region of the smectic phase.
 
However, we have investigated the rich phenomenology of the metastable phase behaviour by restricting the analysis to spatially uniform phases. At high densities, both N$_{\rm r}$-B and N$_{\rm p}$-B are predicted. Depending on the mean value of $\sigma_2$, these transitions may be completed or not. Consequently, the precise shape of the parent size distribution, which is generally difficult to control experimentally, may have a dramatic impact on the phase behaviour observed along dilution paths. In particular, it may prevent the completion of the expected phase separation between uniform phases.

The remainder of the paper is organized as follows. Section \ref{la_teoria} is divided into several subsections. We first define the model (Sec. \ref{model}) and describe how continuous polydispersity is introduced in the intermediate particle dimension (Sec. \ref{parent_section}). We then present the theoretical framework used to study the phase behaviour of polydisperse board-like particles (Sec. \ref{theory_section}), with additional technical details provided in Appendix \ref{app1}. The final subsection (Sec. \ref{magnitudes}) introduces the quantities used to characterize the phase behaviour, the orientational order, and the degree of size fractionation between coexisting phases. Section \ref{results} presents the results and is divided into two parts: the theoretical predictions (Sec. \ref{results_theory} and Appendix \ref{app2}) and the MC simulations (Sec. \ref{results_MC}). Finally, the main conclusions are summarized in Sec. \ref{conclusions}.

\section{Theory}
\label{la_teoria}
In this section we define the model, explain how to incorporate the polydispersity in one of the particle lengths, and also introduce the theoretical formalism used to study the phase behavior of a fluid of polydisperse biaxial particles.

\subsection{Model}
\label{model}

\begin{figure}
	\includegraphics[width=3.0in]{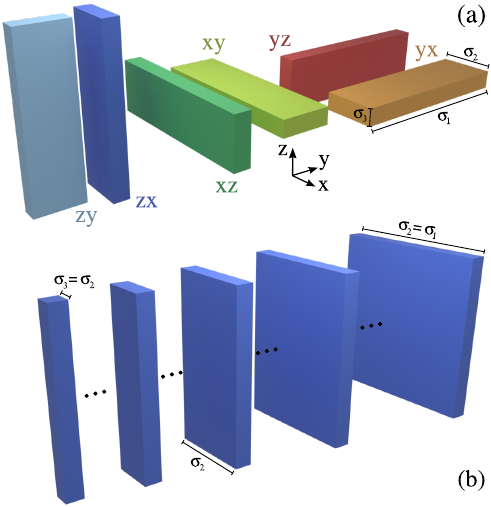}
	\caption{
    (a) Schematic illustration of the particles: biaxial boards with restricted orientations (Zwanzig approximation).
    The distinct particle species (orientations) are colored and labeled according to the alignment of their longest and intermediate edge lengths.
    (b) The boards have their longest and shortest edge lengths fixed to $\sigma_1$ and $\sigma_3$ respectively (with $\sigma_1/\sigma_3=10$ in the figure).
    The intermediate edge length, $\sigma_2$, is polydisperse within the interval $[\sigma_3,\sigma_1]$.
    }
	\label{fig1}
\end{figure}

The model consists on a polydisperse fluid of hard biaxial boards with dimensions $\sigma_1\times\sigma_2\times\sigma_3$, where $\sigma_1\geq \sigma_2\geq \sigma_3$.
While the shortest, $\sigma_3$, and longest, $\sigma_1$, edge lengths are fixed, the intermediate edge length $\sigma_2$ is treated as a polydisperse variable distributed according to a prescribed \emph{parent} probability distribution function, to be defined in Sec.~\ref{parent_section}.
To simplify the treatment of the orientational degrees of freedom, we adopt the Zwanzig model, in which the particle axes are restricted to align with the Cartesian axes.
Therefore, particles can adopt only six distinct orientations, corresponding to the possible permutations of their three principal axes along the $x$, $y$, and $z$ directions.

To implement this approach, it is convenient to introduce the edge length tensor $\sigma_{\mu\nu}^{\tau}$, which specifies the particle dimension of the orientational species $\mu\nu$ along the Cartesian direction $\tau=\{x,y,z\}$. The longest particle axis, of length $\sigma_1$, is oriented along the direction $\mu=\{x,y,z\}$, whereas the intermediate axis, of length $\sigma_2$, is oriented along $\nu=\{x,y,z\}$.
Since $\mu\neq\nu$, there are six distinct species. 
A schematic representation of the polydisperse boards in the Zwanzig approximation is shown in Fig.\ref{fig1}.

The expression for the edge-length tensor is  
\begin{eqnarray}
	\sigma_{\mu\nu}^{\tau}=\sigma_3+\left(\sigma_1-\sigma_3\right)\delta_{\tau\mu}+
	\left(\sigma_2-\sigma_3\right)\delta_{\tau\nu},
\end{eqnarray}
where $\delta_{\alpha\beta}$ is the Kronecker delta.
This tensor will be used in Sec. \ref{theory_section}, where the theoretical formalism is introduced.
Throughout this work, $\sigma_3$ is used as the unit of length and therefore set to unity, while the longest edge length $\sigma_1$ is fixed at either 5 or 10.

\subsection{Parent distribution function}
\label{parent_section}

For the parent probability density function, we use a truncated Schulz distribution,
\begin{eqnarray}
        f(\sigma_2)={\cal C} \left(\frac{\sigma_2}{\sigma_0}\right)^{\nu} e^{-\alpha\sigma_2/\sigma_0},\quad
        \sigma_3\leq \sigma_2\leq \sigma_1, \label{schultz}
\end{eqnarray}
where the normalization constant ${\cal C}$ is determined from the normalization condition $\int_{\sigma_3}^{\sigma_1} d\sigma_2 f(\sigma_2)=1$. The parameter 
$\sigma_0$ is chosen to coincide with the mean intermediate edge length, i.e. 
\begin{eqnarray}
\langle \sigma_2\rangle\equiv \int_{\sigma_3}^{\sigma_1}d\sigma_2\sigma_2f(\sigma_2)=\sigma_0. 
	\label{mean}
\end{eqnarray}

The parameters $\nu$ and $\alpha$, which characterize the shape of the distribution, are obtained by imposing Eqn.~(\ref{mean}), together with a prescribed value of the polydisperse coefficient $s$.
The latter is defined as the relative standard deviation,
\begin{eqnarray}
	s=\sqrt{\frac{\langle \sigma^2_2\rangle}{\sigma_0^2}-1}, \quad 
	\langle \sigma_2^2 \rangle \equiv \int_{\sigma_3}^{\sigma_1}d\sigma_2 \sigma_2^2 f(\sigma_2).
\end{eqnarray}

For $\sigma_1=10$, we consider five values of the polydispersity coefficient, 
$s=\{0.11,\ 0.2,\ 0.3,\ 0.4,\ 0.49\}$, and 
calculate the corresponding phase diagrams. For $\sigma_1=5$ we consider the single value $s=0.42$.

\subsection{Theoretical formalism}
\label{theory_section}
We use one of the most successful formulations of density-functional theory, namely fundamental-measure theory 
(FMT) \cite{M-R1,M-R5}, to study the phase behavior 
of polydisperse hard biaxial boards within the Zwanzig approximation. Within this formalism, the excess part of the free-energy density, $\Phi_{\rm exc}$, depends
on a set of weighted densities $n_{\alpha}({\bm r})$. These are obtained by summing over the orientational species and integrating over the polydisperse variable $\sigma_2$, 
the spatial convolutions of the density 
profiles $\rho_{\mu\nu}({\bm r},\sigma_2)$ with a set of one-body weight functions 
$\omega_{\mu\nu}^{(\alpha)}({\bm r},\sigma_2)$. Note that both the density profiles and the weight functions depend explicitly on the polydisperse variable $\sigma_2$. 

In the uniform limit, in which the number densities of all orientational species are spatially uniform, the weighted densities reduce to 
\begin{eqnarray}
	&&n_{\alpha}= \sum_{\mu,\nu} \int_{\sigma_3}^{\sigma_1} 
	d\sigma_2 \rho_{\mu\nu}(\sigma_2) w_{\mu\nu}^{(\alpha)}(\sigma_2), \label{las_n}\\ 
	&&w_{\mu\nu}^{(\alpha)}(\sigma_2)
	\equiv \int d{\bm r}\omega_{\mu\nu}^{(\alpha)}
	({\bm r},\sigma_2), 
\end{eqnarray}
where $w_{\mu\nu}^{(\alpha)}(\sigma_2)$ are the so-called fundamental measures of species $\mu\nu$. They correspond to the total particle volume ($\alpha=3$), the edge-lengths along the different Cartesian directions ($\alpha=1\tau,  \ \tau=\{x,y,z\}$), and  
the corresponding particle surface areas in different directions 
($\alpha=2\tau$), while for $\alpha=0$ the fundamental measure is simply unity. Explicitly,
\begin{eqnarray}
	&&w_{\mu\nu}^{(0)}(\sigma_2)=1,\quad w_{\mu\nu}^{(3)}(\sigma_2)=\sigma_1\sigma_2\sigma_3, \label{primera}\\
	&&w_{\mu\nu}^{(1\tau)}(\sigma_2)=\sigma_2 \delta_{\nu\tau}+(1-\delta_{\nu\tau})\sigma_{\mu\nu}^{\tau},\\
	&&w_{\mu\nu}^{(2\tau)}(\sigma_2)=\sigma_1\sigma_3\left(\delta_{\nu\tau}+
	(1-\delta_{\nu\tau})\frac{\sigma_2}{\sigma_{\mu\nu}^{\tau}}\right). \label{segunda}
\end{eqnarray}
Carrying out the integration over $\sigma_2$ in Eqn.~(\ref{las_n}), the weighted densities can be expressed as 
\begin{eqnarray}
	&&n_0=\sum_{\mu,\nu} m_{\mu\nu}^{(0)},\quad n_3=\sigma_1\sigma_3\sum_{\mu,\nu} m_{\mu\nu}^{(1)}, \label{la_uno}\\
	&&n_{1\tau}=\sum_{\mu} m_{\mu\tau}^{(1)}+\sum_{\nu\neq \tau,\mu} m_{\mu\nu}^{(0)}\sigma_{\mu\nu}^{\tau},
	\quad \tau=\{x,y,z\}, \label{la_dos}\\
	&&n_{2\tau}=\sigma_1\sigma_3 \left(\sum_{\mu}m_{\mu\tau}^{(0)}+\sum_{\nu\neq \tau,\mu} 
	\frac{m_{\mu\nu}^{(1)}}{\sigma_{\mu\nu}^{\tau}}\right), \quad \tau=\{x,y,z\}.\nonumber\\
    &&\label{la_tres}
\end{eqnarray}
Here, we have introduced the zeroth and first moments of the density distribution functions 
$\rho_{\mu\nu}(\sigma_2)$,
\begin{eqnarray}
	m_{\mu\nu}^{(i)}\equiv \int_{\sigma_3}^{\sigma_1} d\sigma_2\sigma_2^i \rho_{\mu\nu}(\sigma_2), 
	\quad i=0,1. \label{momentos}
\end{eqnarray}
Note that, in Eqns.~(\ref{la_uno})-(\ref{la_tres}), and also in all equations below in this section, all sums are understood not to contain diagonal terms of any quantity.  

Once the weighted densities $n_{\alpha}$ have been expressed in terms of moments of $\rho_{\mu\nu}(\sigma_2)$, the uniform-limit excess free-energy density can be written, within FMT \cite{M-R1}, as
\begin{eqnarray}
	&&\Phi_{\rm exc}\equiv \frac{\beta {\cal F}_{\rm exc}[\{\rho_{\mu\nu}\}]}{V}=\nonumber\\
	&&-n_0 \ln(1-n_3)+\frac{\sum_{\tau} n_{1\tau}n_{2\tau}}{1-n_3}+\frac{\prod_{\tau}n_{2\tau}}{(1-n_3)^2}, \label{excess}
\end{eqnarray}
where ${\cal F}_{\rm exc}[\{\rho_{\mu\nu}\}]$ is the excess part of the Helmholtz density functional, $V$ 
is the system volume, and $\beta^{-1}=k_BT$ is the thermal energy, with $k_B$ the Boltzmann constant and $T$ the temperature.

The ideal part of the free-energy density is given exactly by
\begin{eqnarray}
	&&\Phi_{\rm id}\equiv \frac{\beta {\cal F}_{\rm id}[\{\rho_{\mu\nu}\}]}{V}=\nonumber\\ 
	&&\sum_{\mu,\nu} \int_{\sigma_3}^{\sigma_1} d\sigma_2 \rho_{\mu\nu}(\sigma_2)\left[ 
	\ln \left(\rho_{\mu\nu}(\sigma_2)\right)-1\right],
\end{eqnarray}
where thermal-volume factors have been omitted.

For a single equilibrium phase of a polydisperse fluid, the parent probability distribution function $f(\sigma_2)$ is prescribed. The number-density distributions $\{\rho_{\mu\nu}(\sigma_2)\}$ must therefore satisfy the constraint 
\begin{eqnarray}
	\rho_0 f(\sigma_2)=\sum_{\mu,\nu} \rho_{\mu\nu}(\sigma_2), \label{constraint}
\end{eqnarray}
where $\rho_0$ is the total parent number density. Minimization of the total free-energy density, $\Phi= 
\Phi_{\rm id}+\Phi_{\rm exc}$, with respect to $\rho_{\mu\nu}(\sigma_2)$, subject to the constraint (\ref{constraint}), yields 
\begin{eqnarray}
	\rho_{\mu\nu}(\sigma_2)=\rho_0 f(\sigma_2) \frac{e^{c_{\mu\nu}(\sigma_2)}}
	{\sum_{\tau, \iota} e^{c_{\tau \iota}(\sigma_2)}}, \label{densities}
\end{eqnarray}
where the one-body direct correlation function in the uniform limit is defined through
\begin{eqnarray}
	&&-c_{\mu\nu}(\sigma_2)=\frac{\delta \Phi_{\rm exc}}{\delta \rho_{\mu\nu}(\sigma_2)}
	=\sum_{\alpha} \frac{\partial \Phi_{\rm exc}}{\partial n_{\alpha}} w_{\mu\nu}^{(\alpha)}(\sigma_2), 
	\\
	&&\alpha=\{0, \ 1\tau, \ 2\tau, \ 3\}. \nonumber \label{final1}
\end{eqnarray}

Taking into account that the weighted densities $n_{\alpha}$ depend on the moments $m_{\mu\nu}^{(i)}$, Eqns. (\ref{momentos}), (\ref{constraint}) and (\ref{densities}) yield 
\begin{eqnarray}
	m_{\mu\nu}^{(i)}=\rho_0 \int_{\sigma_3}^{\sigma_1} d\sigma_2 f(\sigma_2) \sigma_2^i 
	\frac{e^{c_{\mu\nu}(\sigma_2)}}{\sum_{\tau,\iota} e^{c_{\tau \iota}(\sigma_2)}},
	\label{set}
\end{eqnarray}
which constitutes a set of self-consistent integral equations for the equilibrium moments $\{m_{\mu\nu}^{(i)}\}$.
These equations were solved iteratively using a mixed fixed-point algorithm with a convergence tolerance of $10^{-10}$. The one-dimensional integrals were evaluated using a globally adaptive quadrature scheme based on Gauss-Kronrod rules, with an absolute error tolerance of $10^{-10}$.

To determine the equilibrium moments $m_{\mu\nu}^{(i,\alpha)}$ of two coexisting phases (where $\alpha$ labels the phase), the constraint (\ref{constraint}) must be replaced by the lever rule,
\begin{eqnarray}
	\rho_0 f(\sigma_2)=\sum_{\alpha} t_{\alpha} \sum_{\mu,\nu} \rho_{\mu\nu}^{(\alpha)}(\sigma_2),
	\label{constraint2}
\end{eqnarray}
where $t_{\alpha}$ is the fraction of the total volume occupied by phase $\alpha$, with $\sum_{\alpha} t_{\alpha}=1$. 

Constrained minimization of the total free energy subject to Eqn.~(\ref{constraint2}) then gives
\begin{eqnarray}
	m_{\mu\nu}^{(i,\alpha)}=\rho_0 \int_{\sigma_3}^{\sigma_1} d\sigma_2 f(\sigma_2) \sigma_2^i 
	\frac{e^{c_{\mu\nu}^{(\alpha)}(\sigma_2)}}
	{\sum_{\theta} t_{\theta} \sum_{\tau,\iota} e^{c_{\tau\iota}^{(\theta)}(\sigma_2)}},
	\label{self}
\end{eqnarray}
which provides a set of coupled self-consistent equations for the moments of the coexisting phases.
These equations should be supplemented by the condition of mechanical equilibrium, namely equality 
of the pressures of any two coexisting phases $\alpha_1$ and $\alpha_2$,
\begin{eqnarray}
	p\left(\rho_0,\{m_{\mu,\nu}^{(i,\alpha_1)}\}\right)=
	p\left(\rho_0,\{m_{\mu,\nu}^{(i,\alpha_2)}\}\right).
	\label{mechanical}
\end{eqnarray}
The pressure is obtained from 
\begin{eqnarray}
\beta p=\frac{\partial \Phi_{\rm exc}}{\partial n_3}.
\end{eqnarray}
From the solution of Eqns.~(\ref{self}) and (\ref{mechanical}), we obtain the equilibrium values of the moments and the 
parent number density $\rho_0$ at two-phase coexistence. The self-consistent equations for the moments were solved
using the same numerical techniques described above, while
the Newton-Raphson method was employed to find $\rho_0$ from the condition of equal pressures.

Note that equality of chemical potentials of all species between 
coexisting phases is ensured by the Lagrange multiplier associated with
the constraint (\ref{constraint2}),
\begin{eqnarray}
\lambda(\sigma_2)=\ln \left[\frac{\rho_0 f(\sigma_2)}{\sum_{\theta} t_{\theta}\sum_{\tau,\iota}e^{c_{\tau\iota}^{(\alpha)}(\sigma_2)}
}\right].
\end{eqnarray}
The chemical potential associated with particles of size $\sigma_2$ is therefore given by $\lambda(\sigma_2)$, independent of their orientational species $\mu\nu$, and takes the same value in all coexisting phases.

For second-order transitions, either between two uniform phases or between a uniform and a nonuniform phase, the corresponding spinodal was determined using a formalism analogous to the condition of a vanishing inverse structure factor, generalized to polydisperse systems. In the present case, this condition reduces to finding the zeros of the determinant of an $8\times8$ matrix. Details of the formalism and of the numerical procedure used to determine the spinodal are given in Appendix \ref{app1}.

\subsection{Characterization of phases with different symmetries and compositions}
\label{magnitudes}
The parent distribution function is characterized by its mean intermediate edge length $\sigma_0$ and its total packing fraction $\eta_0=\rho_0 \sigma_1\sigma_0 \sigma_3$.
To quantify fractionation between coexisting, phases we define the probability distribution for the intermediate edge length in phase $\alpha$ as
\begin{eqnarray}
	&&f_{\alpha}(\sigma_2)=\frac{\sum_{\mu,\nu}\rho_{\mu\nu}^{(\alpha)}(\sigma_2)}{m_{\alpha}^{(0)}}, \\
	&&m_{\alpha}^{(0)}\equiv \sum_{\mu,\nu} m_{\mu\nu}^{(0,\alpha)}
	\Rightarrow \int_{\sigma_3}^{\sigma_1} d\sigma_2 f_{\alpha}(\sigma_2)=1,
\end{eqnarray}
where $m_{\alpha}^{(0)}$ is the total zeroth moment, and hence the total number density, of phase $\alpha$. The mean intermediate edge length in the coexisting phase $\alpha$ can then be calculated as
\begin{eqnarray}
	\langle \sigma_2\rangle_{\alpha}\equiv \int_{\sigma_3}^{\sigma_1} d\sigma_2\sigma_2 f_{\alpha}(\sigma_2).
\end{eqnarray}
The total packing fraction of phase $\alpha$ is 
\begin{eqnarray}
	\eta_{\alpha}=\sigma_1\sigma_3 m_{\alpha}^{(1)}=\sigma_1\sigma_3\sum_{\mu,\nu} m_{\mu\nu}^{(1,\alpha)},
\end{eqnarray}

To characterize the orientational order of a given phase, we first define the  orientational 
fractions as
\begin{eqnarray}
	\gamma_{\mu\nu}\equiv \frac{m_{\mu\nu}^{(0)}}{m^{(0)}}=\frac{m_{\mu\nu}^{(0)}}{\sum_{\tau,\iota} m_{\tau\iota}^{(0)}}.
\end{eqnarray}
In the isotropic phase, all six 
orientations are equally populated, so that $\gamma_{\mu\nu}=\frac{1}{6}$. 

It is useful to distinguish between rod-like and plate-like uniaxial nematic order. For the rod-like nematic phase, taking the nematic director along the $z$-axis, the  fractions 
satisfy the symmetry relations $\gamma_{zx}=\gamma_{zy}$, $\gamma_{yz}=\gamma_{xz}$ and $\gamma_{xy}=\gamma_{yx}$. We therefore define the rod-like uniaxial 
nematic order parameter as 
\begin{eqnarray}
	Q_{\rm r}=\frac{1}{2}\left(3(\gamma_{zx}+\gamma_{zy})-1\right).
\end{eqnarray}
For the plate-like uniaxial nematic phase, again taking the nematic director along $z$, the symmetry relations are again
$\gamma_{xy}=\gamma_{yx}$, $\gamma_{yz}=\gamma_{xz}$ and $\gamma_{zx}=\gamma_{zy}$. The plate-like uniaxial nematic order parameter in then defined as 
\begin{eqnarray}
	Q_{\rm p}=\frac{1}{2}\left(3(\gamma_{xy}+\gamma_{yx})-1\right).
\end{eqnarray}

Note that the
three pairwise symmetry relations mentioned above are actually identical for the rod-like and plate-like nematics. What distinguishes the two nematics is which pair of orientations is preferentially populated.
For rods, $\gamma_{zx}+\gamma_{zy}$ measures the fraction of particles whose longest axis ($\mu=z$) lies along the director. Hence $Q_{\rm r}\to 1$ for perfect rod-like alignment. For plates, $\gamma_{xy}+\gamma_{yx}$ measures the fraction whose shortest axis lies along $z$: since neither the longest nor intermediate axis is along $z$, it must be the remaining (short) axis. Thus $Q_{\rm p}\to 1$ for perfect plate-like alignment.

In the biaxial nematic phase, the orientational fractions are, in general, all different. We define the biaxial order parameters for rod-like and plate-like ordering as
\begin{eqnarray}
	\Delta_{\rm r}=\frac{1}{2}\left|\gamma_{yx}-\gamma_{yz}+\gamma_{xz}-\gamma_{xy}+2(\gamma_{zx}-\gamma_{zy})\right|,\\
	\Delta_{\rm p}=\frac{1}{2}\left|\gamma_{zx}-\gamma_{xz}+\gamma_{yz}-\gamma_{zy}+2(\gamma_{yx}-\gamma_{xy})\right|.
\end{eqnarray}
These biaxial order parameters vanish in the corresponding uniaxial nematic phases and approach unity in the limit of perfect biaxial order. Specifically, $\Delta_{\rm r}=1$ when $\gamma_{zx}=1$ and all other orientational fractions vanish, whereas $\Delta_{\rm p}=1$ when $\gamma_{yx}=1$ and all other fractions vanish.
If the principal nematic director is not aligned with the $z$ axis, the corresponding expressions for $Q_{\alpha}$ and $\Delta_{\alpha}$ are obtained from those given above by applying the appropriate cyclic permutations of the Cartesian axes, $x\to y$, $y\to z$, and $z\to x$.

\section{Results}
\label{results}

\begin{figure*}
        \includegraphics[width=0.99\textwidth]{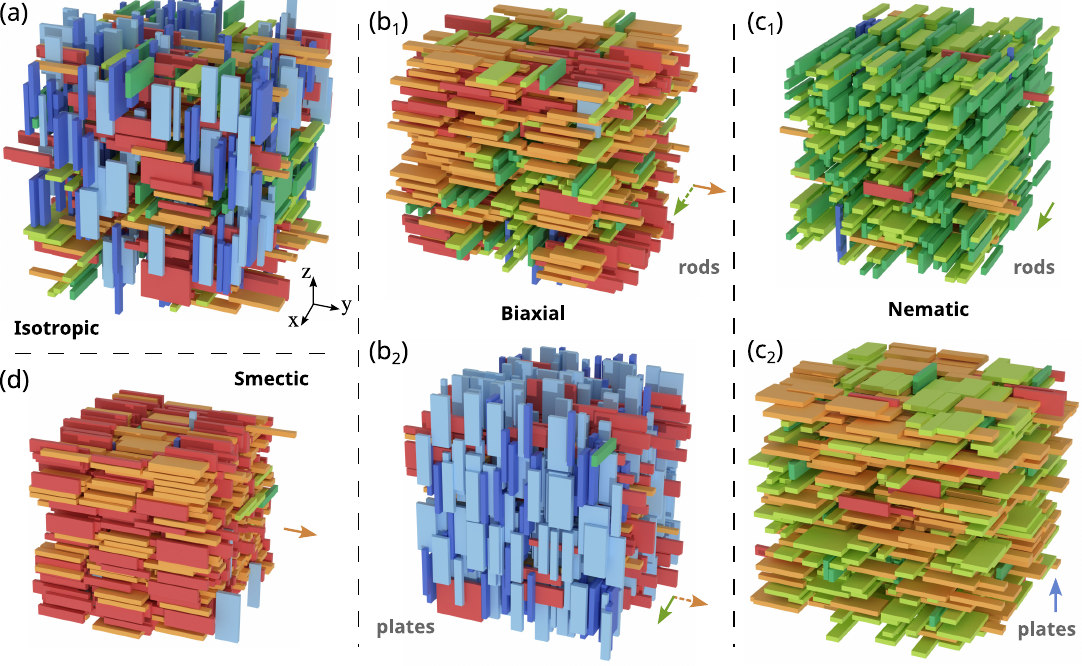}
        \caption{Monte Carlo simulation snapshots of particle configurations ($N=1000$) exhibiting several liquid crystalline phases in a cubic box with periodic boundary conditions.
        (a) I fluid at packing fraction $\eta_0=0.23$ and average intermediate length $\sigma_0=2.4$,
        (b$_1$) B phase of rods ($\eta_0=0.32$ and $\sigma_0=2.4$),
        (b$_2$) B phase of plates ($\eta_0=0.42$ and $\sigma_0=2.7$),
        (c$_1$) N$_{\rm r}$ phase ($\eta_0=0.34$ and $\sigma_0=2.15$),
        (c$_2$) N$_{\rm p}$ phase ($\eta_0=0.39$ and $\sigma_0=3.85$),
        and (d) rod-rich Sm ($\eta_0=0.5$ and $\sigma_0=2.4$).
        Particles are colored according to their orientation, following the color scheme in Fig.~\ref{fig1}.
        The main director is indicated by a solid colored arrow.
        The secondary director in the biaxial phases is indicated by a dashed colored arrow.
        The polydisperse coefficient is $s=0.49$ in all cases.
        }
        \label{fig2}
\end{figure*}

\subsection{Theoretical results}
\label{results_theory}

We first briefly describe the different phases that can be stabilized within the Zwanzig model of polydisperse biaxial boards. The orientationally most disordered phase is the isotropic phase (I), in which all particle orientations are equally populated
[see Fig.~\ref{fig2}(a)], i.e. all the  
fractions satisfy $\gamma_{\mu\nu}=\frac{1}{6}$. At
higher densities and sufficiently small values of the mean intermediate edge length, $\sigma_0$, 
the rod-like uniaxial nematic phase (N$_{\rm r}$) can be stabilized. In this phase the longest particle axes are preferentially aligned along the principal nematic director, whereas the intermediate axes are equally distributed between the two equivalent Cartesian directions perpendicular to the director [see Fig.~\ref{fig2}(c$1$)].
Conversely, for sufficiently large values of $\sigma_0$, the plate-like  
uniaxial nematic phase (N$_{\rm p}$) can become stable. In this case, the 
shortest particle axes are preferentially aligned along the principal nematic director, while the intermediate axis is equally distributed  
between the two equivalent perpendicular directions [see Fig.~\ref{fig2}(c$_2$)].
In both uniaxial nematic phases the six orientational species are grouped into three symmetry-related pairs and are therefore characterized by three independent orientational fractions.

When the equivalence between the two directions perpendicular to the principal nematic director is broken in the N$_{\rm r,p}$ phases, the system develops biaxial nematic order, giving rise to the B phase [see Fig.~\ref{fig2}(b)]. In this phase, the orientational fractions are, in general, all different. 

Finally, at sufficiently high packing fractions and small or large values of $\sigma_0$ 
the smectic phase (Sm) can be stabilized, corresponding predominantly to rod-like and plate-like particles, respectively [see Fig.~\ref{fig2}(d)]. The smectic period is associated with the longest particle length (rod-like regime) or with the shortest particle dimension (plate-like regime). The Sm phase may exhibit either uniaxial 
or biaxial orientational ordering depending on the precise value of $\sigma_0$.

\begin{figure*}
\includegraphics[width=2.1in]{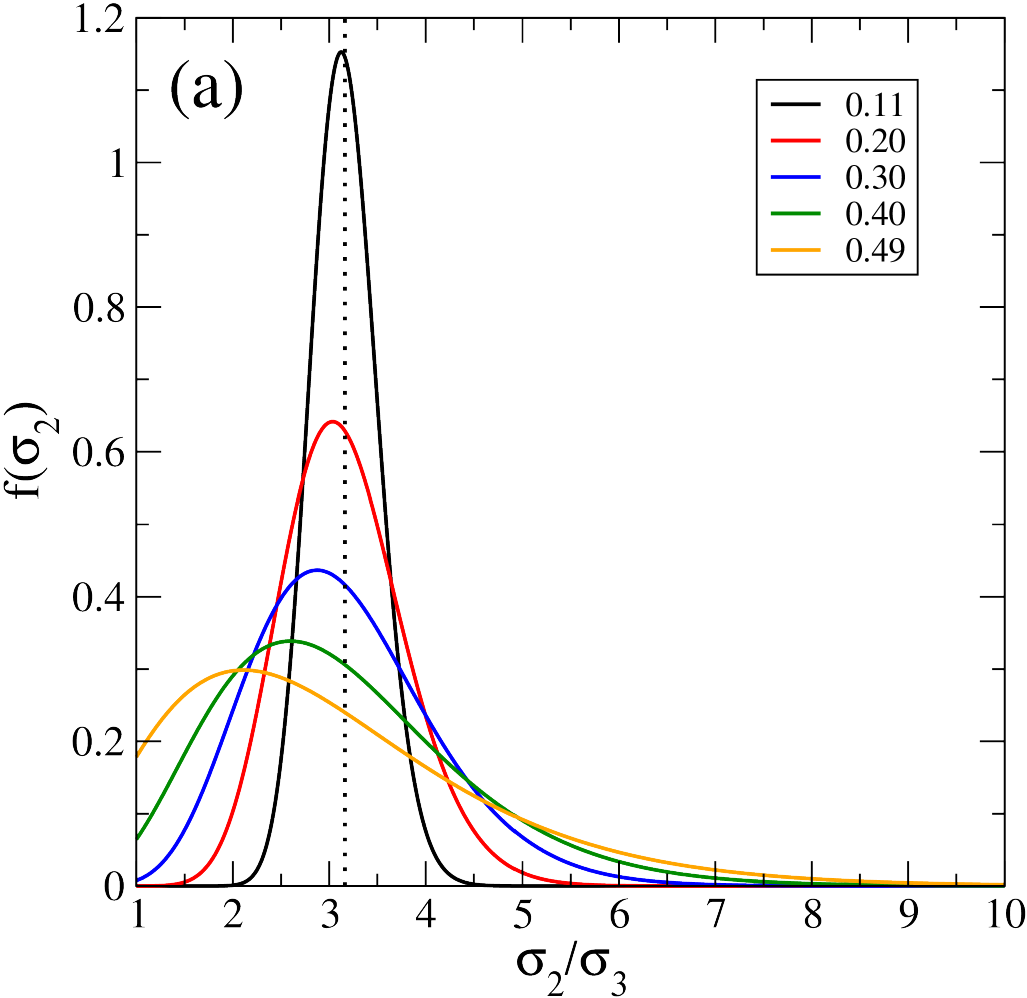}
\includegraphics[width=2.1in]{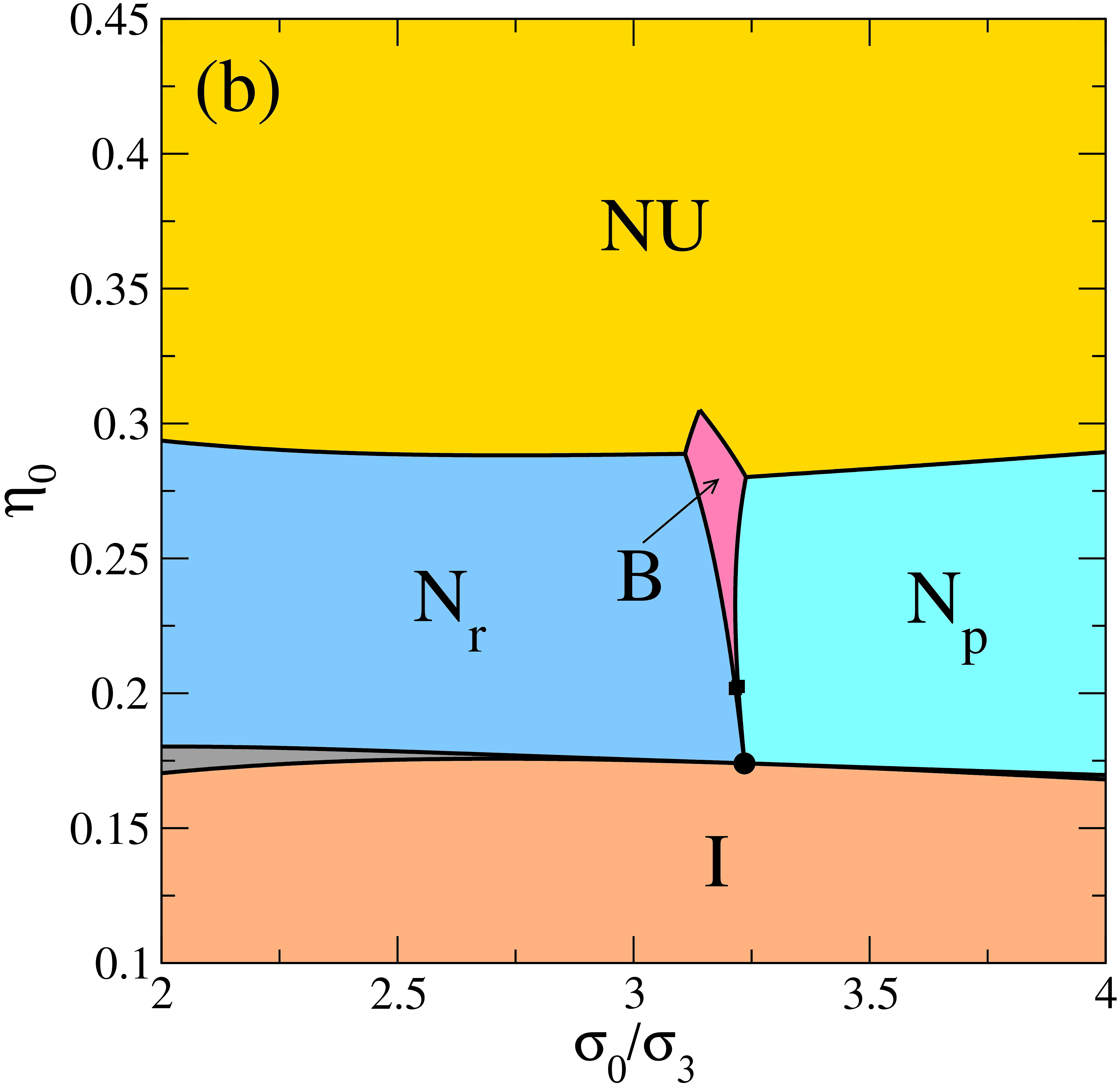}
\includegraphics[width=2.1in]{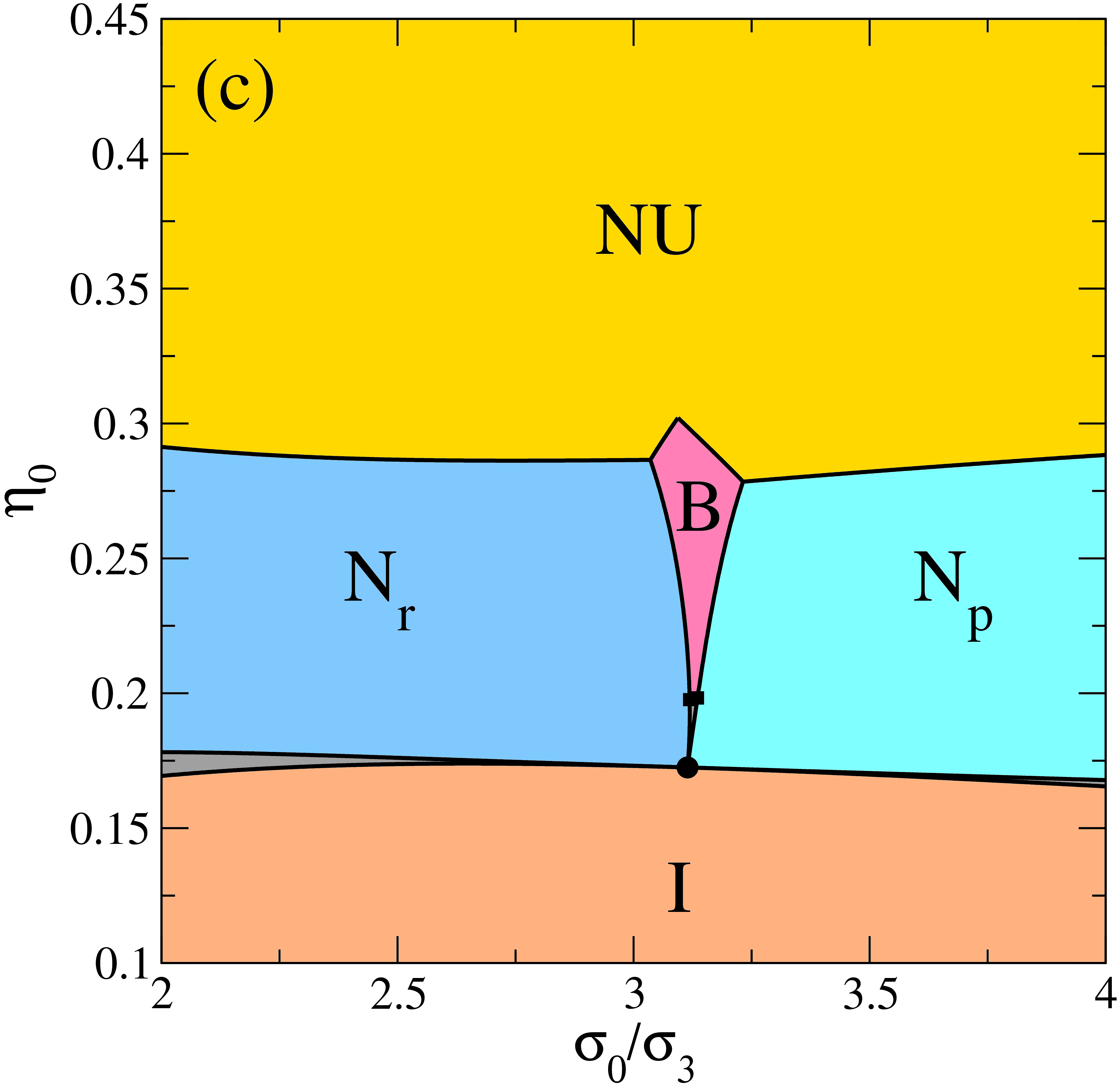}
\includegraphics[width=2.1in]{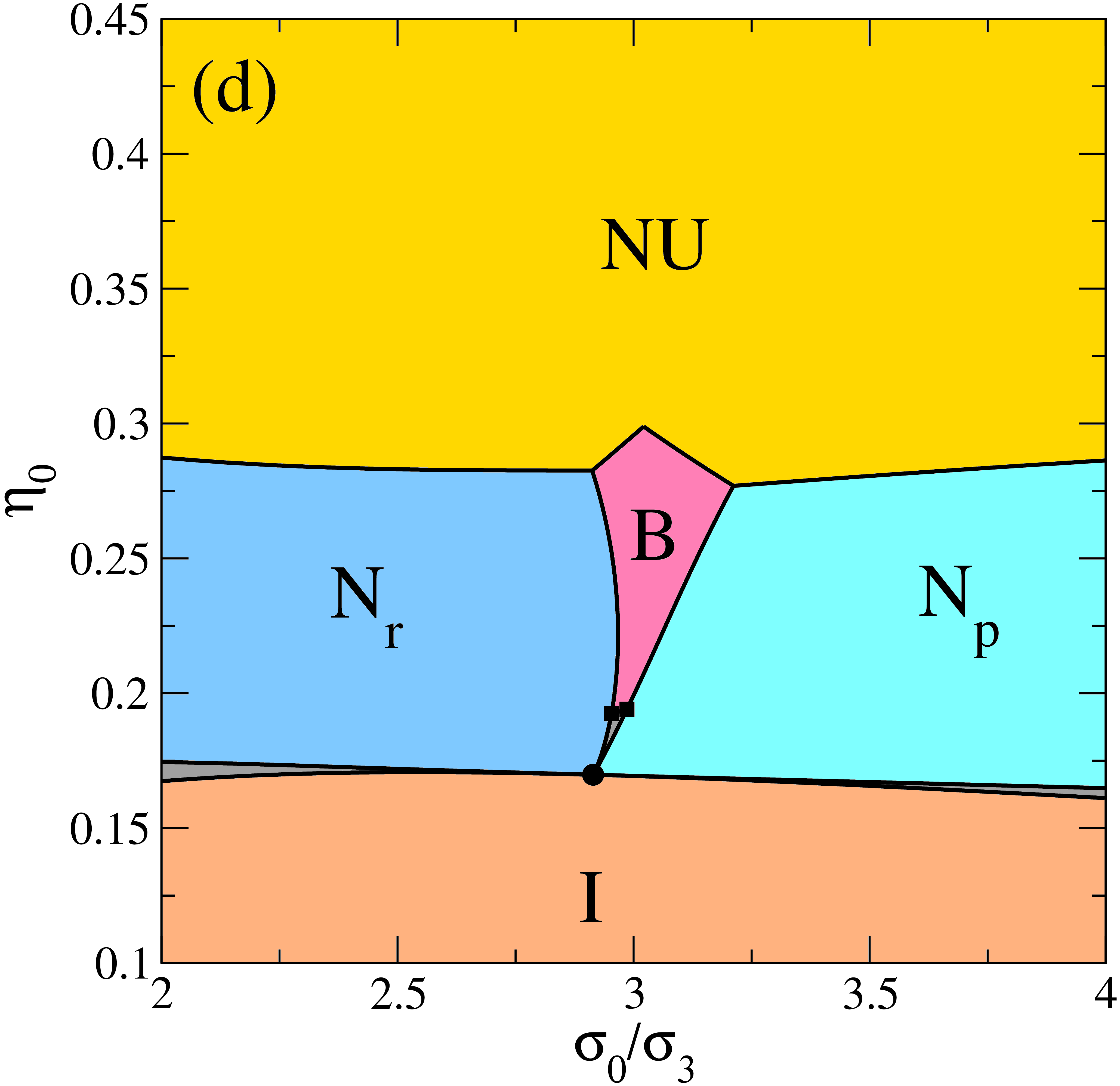}
\includegraphics[width=2.1in]{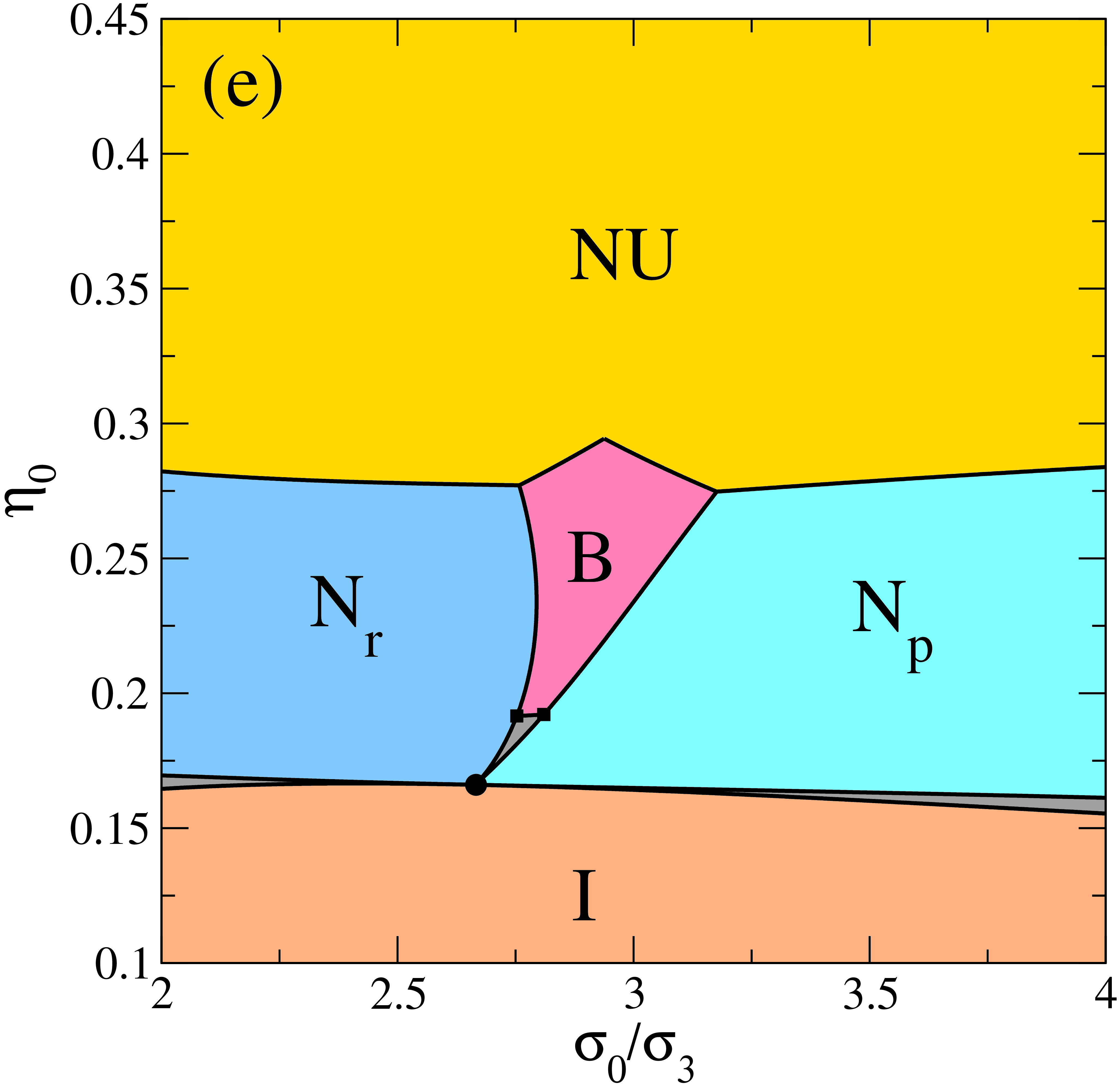}
\includegraphics[width=2.1in]{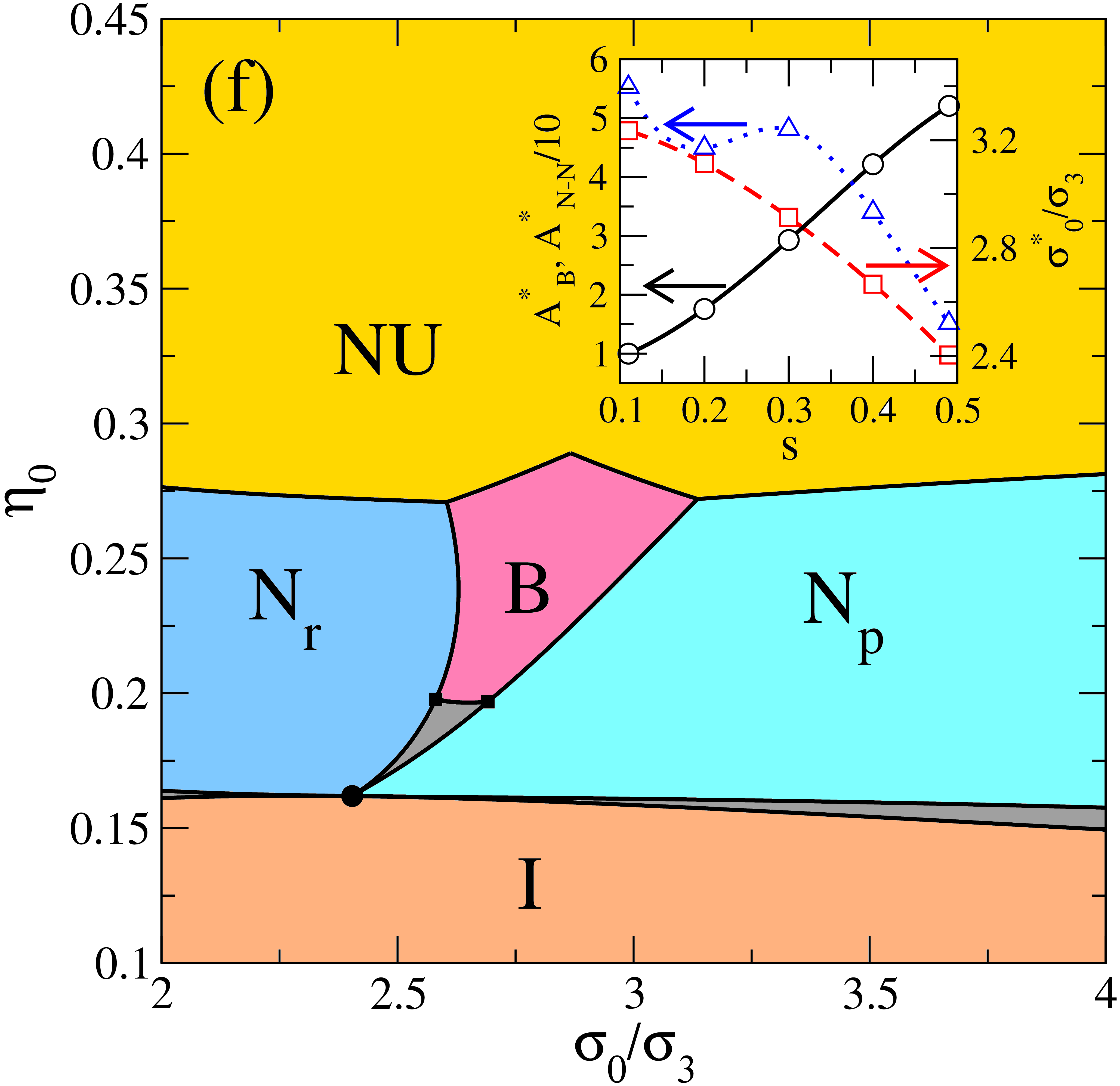}
        \caption{Parent probability density distributions 
        $f(\sigma_2)$ for different values of the polydispersity coefficient $s$
        and a fixed value of the mean intermediate particle length
        $\sigma_0=\sqrt{\sigma_1\sigma_3}$ (vertical dotted line) 
        corresponding to perfectly biaxial boards. (b)-(f) Phase diagrams in the packing fraction, $\eta_0$, vs.
        mean intermediate length $\sigma_0$ of the parent
        number-density distribution function, for polydispersity coefficients $s=0.11$ (b), 0.2 (c), 0.3 (d),
        0.4 (e), and 0.49 (f). The regions of stability of the different phases 
        are labelled accordingly. The gray shaded regions indicate the
        I-N and N$_{\rm r}$-N$_{\rm p}$ coexistence instability regions. 
        In addition to the binodals associated with
        these first-order transitions, second-order N$_{\rm r,p}$--B transition lines are present. At high packing fractions, these lines meet the spinodals associated with the (N$_{\rm r,p}$,B)--Sm instabilities.
Inset of panel (f): The fraction of area of the B-phase stability region, normalized by its value at the lowest polydispersity, $s=0.11$, as a function of $s$ (solid line). Also shown is the area of the N$_{\rm r}$--N$_{\rm p}$ demixing region, normalized by its value at $s=0.11$, as a function of $s$ (dotted line). The location of the tetracritical point, $\eta_0^*$, as a function of $s$ is shown by the dashed line.
        }
        \label{fig3}
\end{figure*}

We now present the main results obtained from the numerical
implementation of the formalism developed in Sec. \ref{theory_section}. In
particular, we have solved Eqns.~(\ref{self}) and (\ref{mechanical}) to determine
two-phase coexistence, and Eqns.~(\ref{bifurca}) and (\ref{necessary}) to determine spinodal
instability conditions. We looked for coexistence between the I
and N$_{\rm r}$ or N$_{\rm p}$ phases, as well as between the two nematic phases, the latter corresponding to a demixing transition.

Regarding the instability conditions, we have determined 
the bifurcation points between uniform phases, i.e. the packing
fraction values at which the N$_{\rm r,p}$ lose stability with respect to the B
phase. We have also calculated
the bifurcations from the N$_{\rm r,p}$ and B phases to nonuniform (NU) phases, in particular to Sm phases.

\begin{figure}[ht]
        \includegraphics[width=3.in]{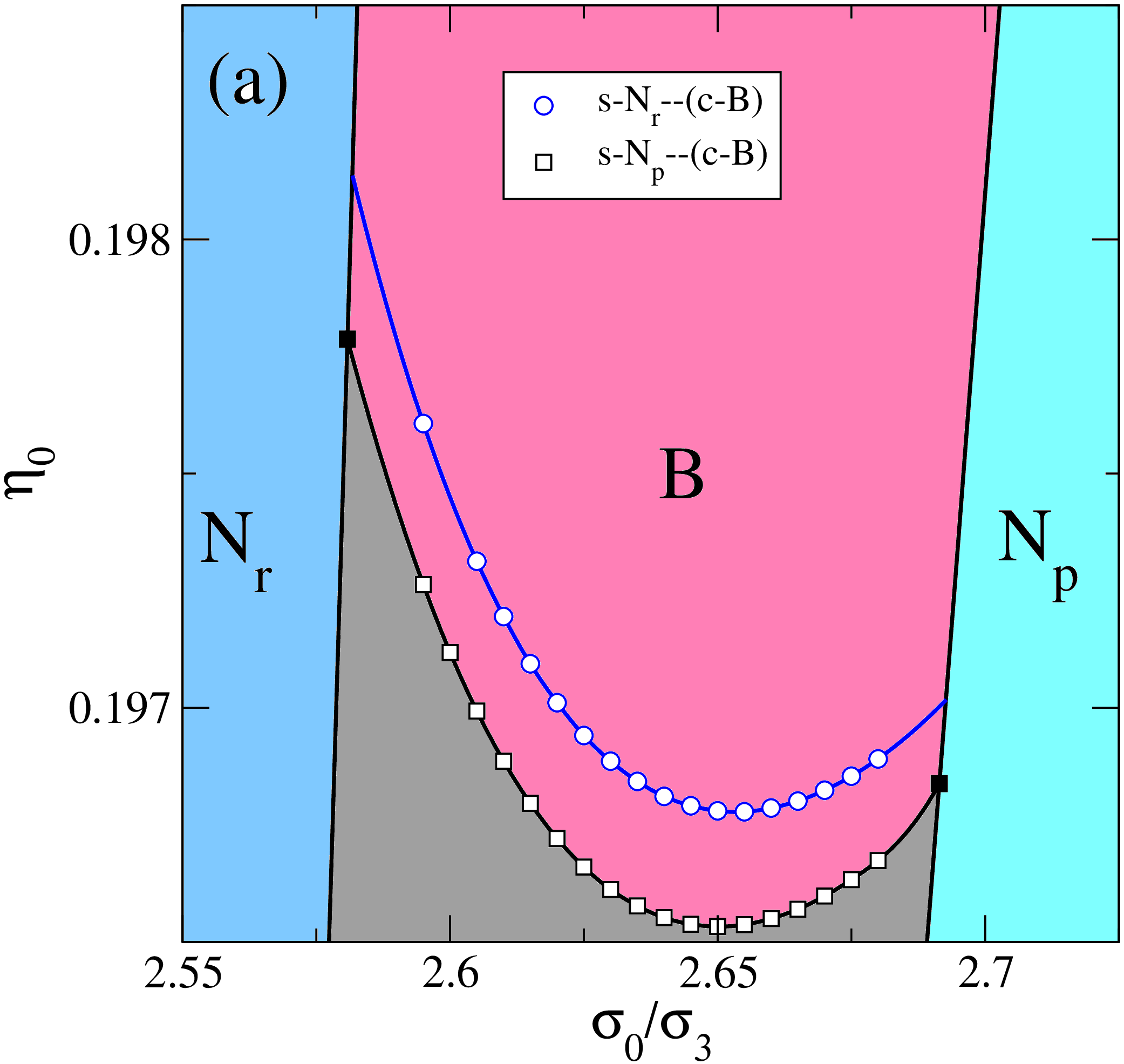}
        \includegraphics[width=3.in]{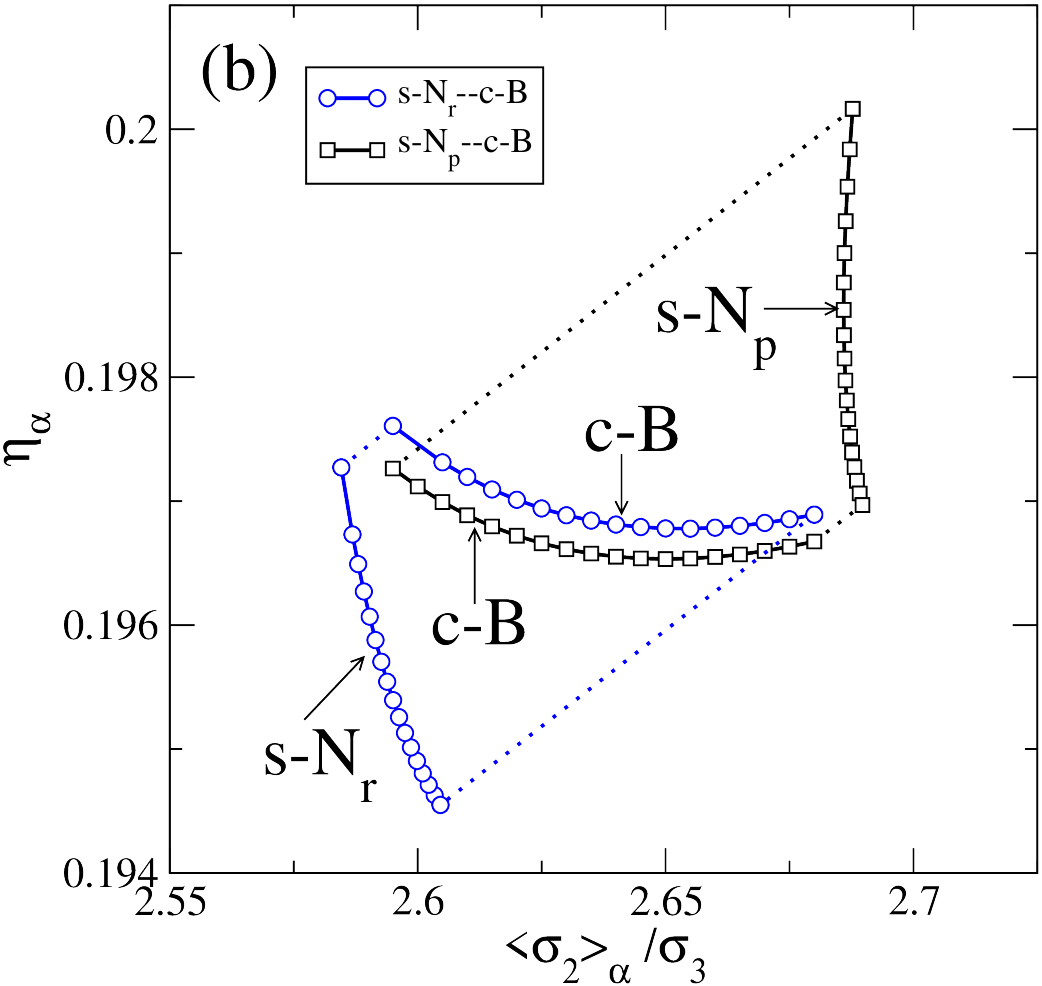}
        \caption{
        (a) Detail of the phase diagram shown in Fig.~\ref{fig3}(f), displaying the first-order B--N$_{\rm r}$ and B--N$_{\rm p}$ transitions, the latter being the stable one. The B phase occupies the entire volume and corresponds to the cloud phase, whereas the nematic phases of rods and plates are the corresponding shadow phases and occupy an infinitesimal fraction of the total volume. (b) Packing fractions $\eta_{\alpha}$ (with $\alpha={\rm N}_{\rm r,p}$) of the coexisting N${\rm r}$ and N$_{\rm p}$ shadow phases, together with $\eta_0$, as functions of the mean intermediate particle length $\langle \sigma_2\rangle_{\alpha}$, calculated using the corresponding shadow-phase distribution functions $f_{\alpha}(\sigma_2)$.}
        \label{fig4}
\end{figure}

For the calculations, we have used as parent distributions $f(\sigma_2)$
the truncated Schulz distributions ($\sigma_2\in[\sigma_3,\sigma_1]=[1,10]$)
 shown in Fig.~\ref{fig1}(a), for the
following fixed values of the polydispersity coefficient:
$s=\{0.11,\ 0.2,\ 0.3, \ 0.4, \ 0.49\}$. For illustrative purposes, we have selected
the mean intermediate length corresponding to perfectly biaxial boards,
$\sigma_0=\sqrt{\sigma_1\sigma_3}$ (vertical line). However, this length is
the variable that we vary to calculate the phase diagrams in the plane defined by the 
packing fraction of the parent phase, $\eta_0$, vs. $\sigma_0$. Note that
this choice of variables implies that only the cloud
coexistence curves are represented, i.e. those corresponding to phases that occupy the entire sample volume and coexist with phases occupying an infinitesimal fraction of it
(the shadow phases). Thus, the cloud and parent phases coincide. 

We vary the mean intermediate length $\sigma_0$ from
2 to 4, so that the predominant particle shape in the polydisperse mixture changes from rod-like to plate-like.
The resulting phase diagrams for the different polydispersities
are shown in Fig.~\ref{fig3} (b)-(f).
As can be seen, the I phase coexists with the N$_{\rm r}$ or
N$_{\rm p}$ phase to the left or right, respectively, of the multicritical point [filled circles in
Fig.~\ref{fig3} (b)-(f)], with the transition becoming progressively weaker as the binodals
approach this point. The N$_{\rm r}$-N$_{\rm p}$ demixing transition emerges from the multicritical point, and its
binodals meet the N$_{\rm r,p}$-B spinodals at the points indicated by squares.

From these points, two second-order transition curves, separating the N$_{\rm r,p}$ and B
phases, emerge and delimit the stability region of the B phase on the left and right. At higher packing fractions, these
curves meet the N$_{\rm r,p}$-Sm spinodal curves. The B phase is bounded from above by the
B-Sm spinodal, which exhibits a cusp. This indicates that, on the two sides of the cusp, the equilibrium smectic period is associated with the largest ($\sigma_1$) or smallest ($\sigma_3$) particle lengths, respectively. From below, the B phase
is bounded by a first-order B-N$_{\rm p}$ transition, as shown in Fig.~\ref{fig4} for the particular case $s=0.49$.

For completeness, we also show both cloud-B--shadow-N$_{\rm r,p}$
coexistence branches [Fig.~\ref{fig4}(b)]. The B-N$_{\rm p}$ occurs at a lower value
of $\eta_0$, indicating that this is the thermodynamically stable transition.

\begin{figure}[!htbp]
        \includegraphics[width=3.0in]{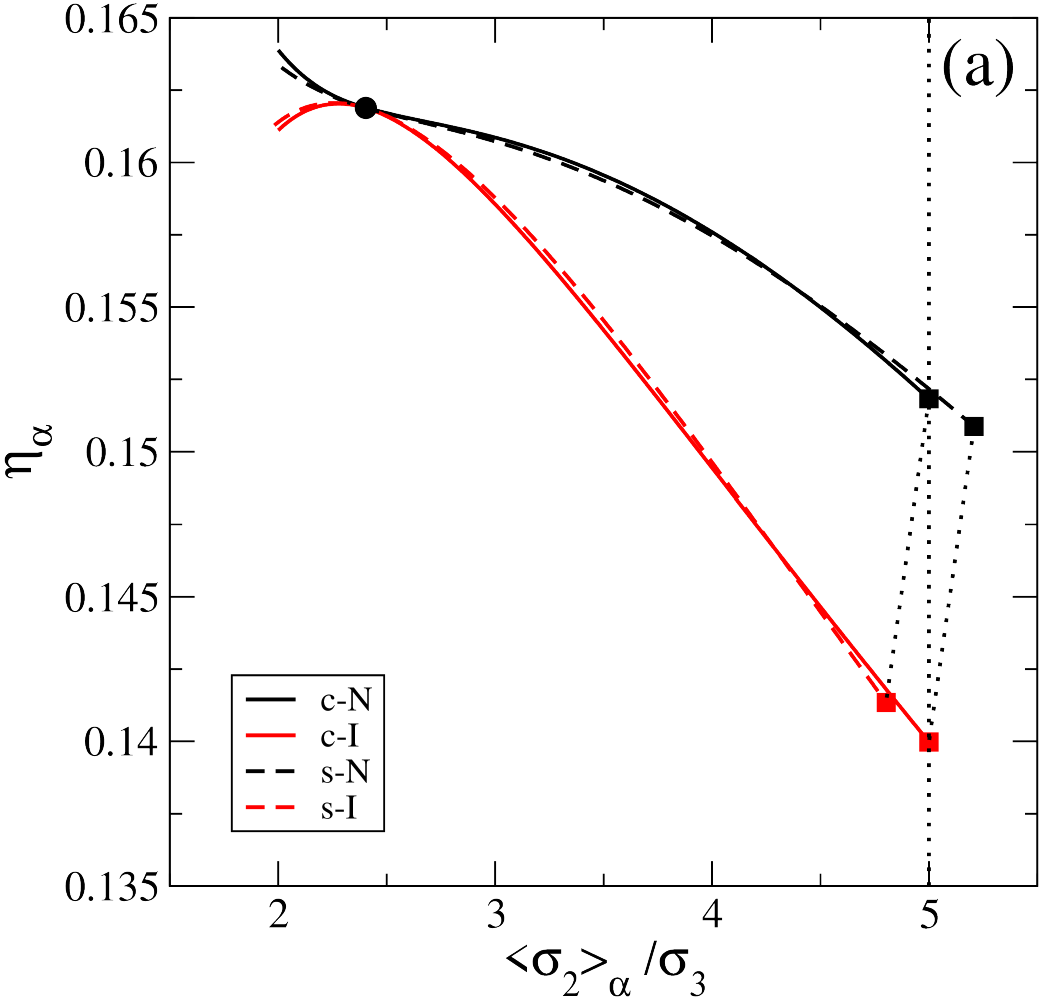}
        \includegraphics[width=3.0in]{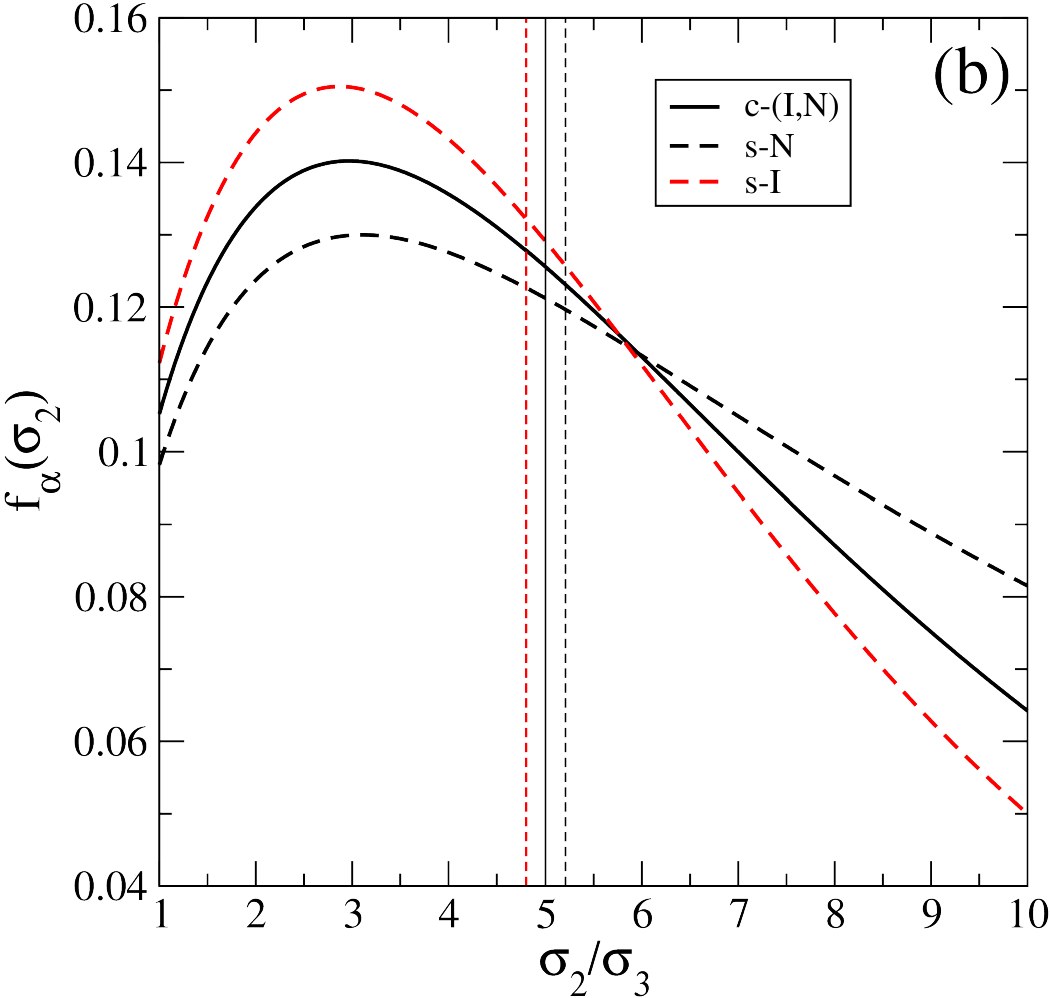}
        \caption{(a) Packing fractions, $\eta_{\alpha}$, along the cloud-I--shadow-N and shadow-I--cloud-N coexistence branches as functions of the mean intermediate length, $\langle\sigma_2\rangle_{\alpha}$, calculated with the corresponding equilibrium distribution functions. The polydispersity coefficient is fixed at $s=0.49$. The coexistence points corresponding to $\sigma_0=5$ (vertical dotted line) are indicated by squares. (b) Coexisting distribution functions $f_{\alpha}(\sigma_2)$ corresponding to $\sigma_0=5$. Vertical lines indicate the mean values $\langle\sigma_2\rangle_{\alpha}$ calculated from the corresponding distributions $f_{\alpha}(\sigma_2)$, revealing a pronounced fractionation of particle sizes between the coexisting phases.}
        \label{fig5}
\end{figure}

As can be seen from Figs.~\ref{fig3}(b)-(f), the multicritical point shifts towards lower values of both $\sigma_0$ and 
$\eta_0$ as the polydispersity increases. This behaviour can be understood in terms of the increasing contribution of plate-like particles to the mixture. Since the plate-like particles have larger volumes, their 
excluded volumes in the isotropic phase are correspondingly larger, favouring orientational ordering into the N$_{\rm p}$ phase.
As a result, the I-N$_{\rm p}$ transition occurs at lower values of $\sigma_0$ and $\eta_0$.  

Also, increasing the polydispersity $s$ enlarges both the N$_{\rm r}$-N$_{\rm p}$ demixing region, and the region of stability of the B phase. However the latter grows more rapidly than the former, allowing us to conclude that 
the polydispersity has an overall stabilizing effect on the B phase. To quantify these trends, the inset of panel (f) shows the 
areas of the N$_{\rm r}$--N$_{\rm p}$ demixing region ($A_{\rm N-N}^*$) and the B-phase stability region ($A_{\rm B}^*$), normalized by their respective values at the lowest polydispersity, $s=0.11$, as functions of $s$. 
We also show the curve $\sigma_0^*$ (the location of the multicritical point) 
as a function of $s$.  

This enhancement of B-phase stability with increasing 
polydispersity was already reported in Ref.~\cite{Patti1}, where a multicomponent 
mixture containing a large fraction of board-like particles was studied theoretically within the Onsager density-functional approximation for the Zwanzig model. In that work, however, 
a constant-composition approximation was used to calculate the two-phase 
coexistence, thereby excluding the possible existence of N$_{\rm r}$-N$_{\rm p}$ demixing.
Here we show that such demixing does indeed occur, but that it is not sufficiently strong, at least for the values of $\sigma_1$ and $s$ explored here, to suppress the stabilizing effect of polydispersity on the B phase.

\begin{figure}[!htbp]
        \includegraphics[width=3.0in]{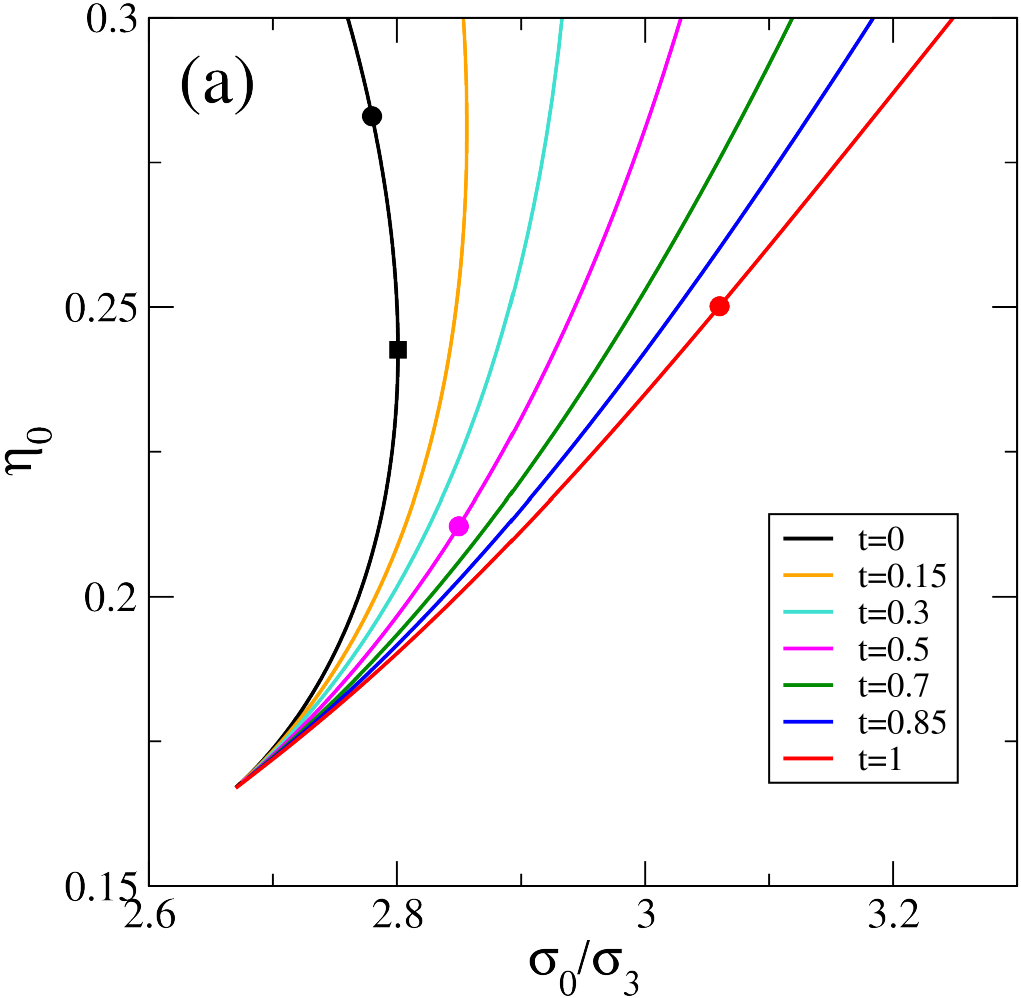}
        \includegraphics[width=3.0in]{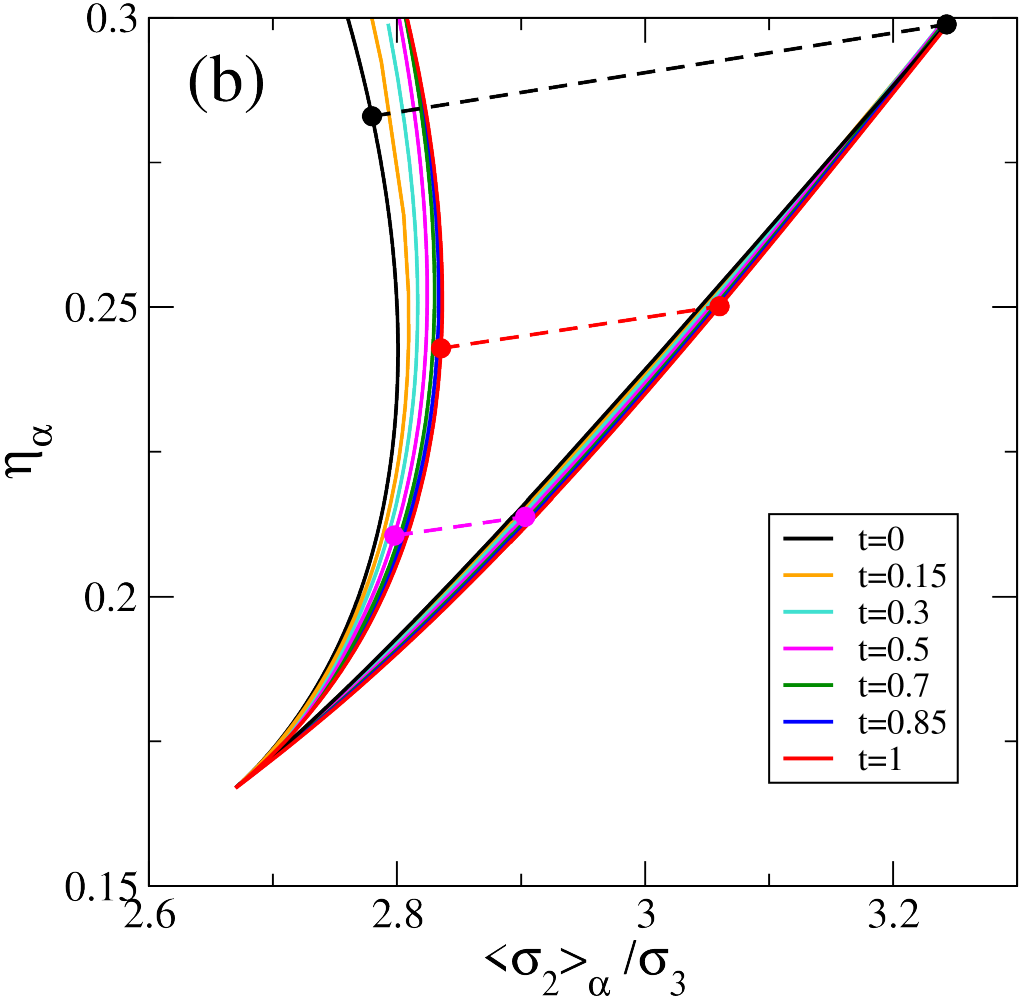}
        \caption{
        (a) Packing fraction of the parent distribution, $\eta_0$, as a function of its mean intermediate length, $\sigma_0$, along the N$_{\rm r}$--N$_{\rm p}$ demixing transition for different values of the volume fraction $t$ occupied by the N$_{\rm p}$ phase, at a fixed polydispersity coefficient $s=0.4$. Three representative cases, corresponding to $t=0$, $0.5$, and $1$, are indicated by solid circles to illustrate the fractionation effect. The solid square indicates the maximum of the curve $\sigma_0(\eta_0)$ for $t=0$. (b) Coexisting binodals represented in terms of the packing fraction and the mean intermediate length, both calculated using the distribution functions $f_{\alpha}(\sigma_2)$ ($\alpha={\rm N}_{\rm r,p}$) of the coexisting phases. Symbols connected by dashed lines indicate pairs of coexisting phases corresponding to the same values of $\eta_0$ and $\sigma_0$ as those indicated in panel (a).}
        \label{fig6}
\end{figure}

One of the main purposes of the present work is to assess the role of the N-N demixing in the phase behavior of polydisperse biaxial boards. As already pointed out,  
a region of N$_{\rm r}$-N$_{\rm p}$ demixing always appears in the phase 
diagram just above the multicritical point, although it is rather
small at low polydispersities. Since polydispersity is introduced through a 
unimodal parent distribution function, 
which is relatively narrow for small values of $s\approx 0.1$, 
it is remarkable that N-N demixing persists even in this regime and occupies a finite region of the phase diagram.

More importantly, coexistence is accompanied by fractionation 
of particles with different shapes between the two coexisting phases. This effect 
is also present in the I-N$_{\rm r,p}$ transition as illustrated in 
Fig.~\ref{fig5}. In panel (a) we plot the coexisting packing fractions, 
$\eta_{\rm I,N}$ and mean intermediate lengths, $\langle \sigma_2\rangle_{\rm I,N}$, 
calculated with the corresponding equilibrium length-distribution functions 
$f_{\rm I,N}(\sigma_2)$. Only the cloud (shadow)-I-- shadow (cloud)-N 
coexistence branches are shown. The cloud and shadow curves corresponding to the same phase lie close to 
each other. Fractionation becomes more apparent when coexisting points are compared, as illustrated in panel (a) for $\sigma_0=5$.

Note that the I phase, whether cloud or shadow, always has a lower value of the mean intermediate length than the coexisting N$_{\rm p}$ (cloud or shadow) phase, indicating that the latter is enriched in boards with a more 
pronounced plate-like geometry. This behaviour is confirmed by the 
coexisting distribution functions $f_{\alpha}(\sigma_2)$ shown in panel (b). The distribution corresponding to the I phase has a higher maximum and decays more rapidly at large $\sigma_2$ than that of the coexisting N$_{\rm p}$ phase. 
The significantly slower decay of $f_{{\rm N}_{\rm p}}(\sigma_2)$  at 
large $\sigma_2$ reflects the preferential partitioning of the more plate-like particles into the N$_{\rm p}$ phase, resulting in a larger mean value $\langle \sigma_2\rangle_{\rm N_{\rm p}}$.

\begin{figure*}[!htbp]
        \includegraphics[width=2.3in]{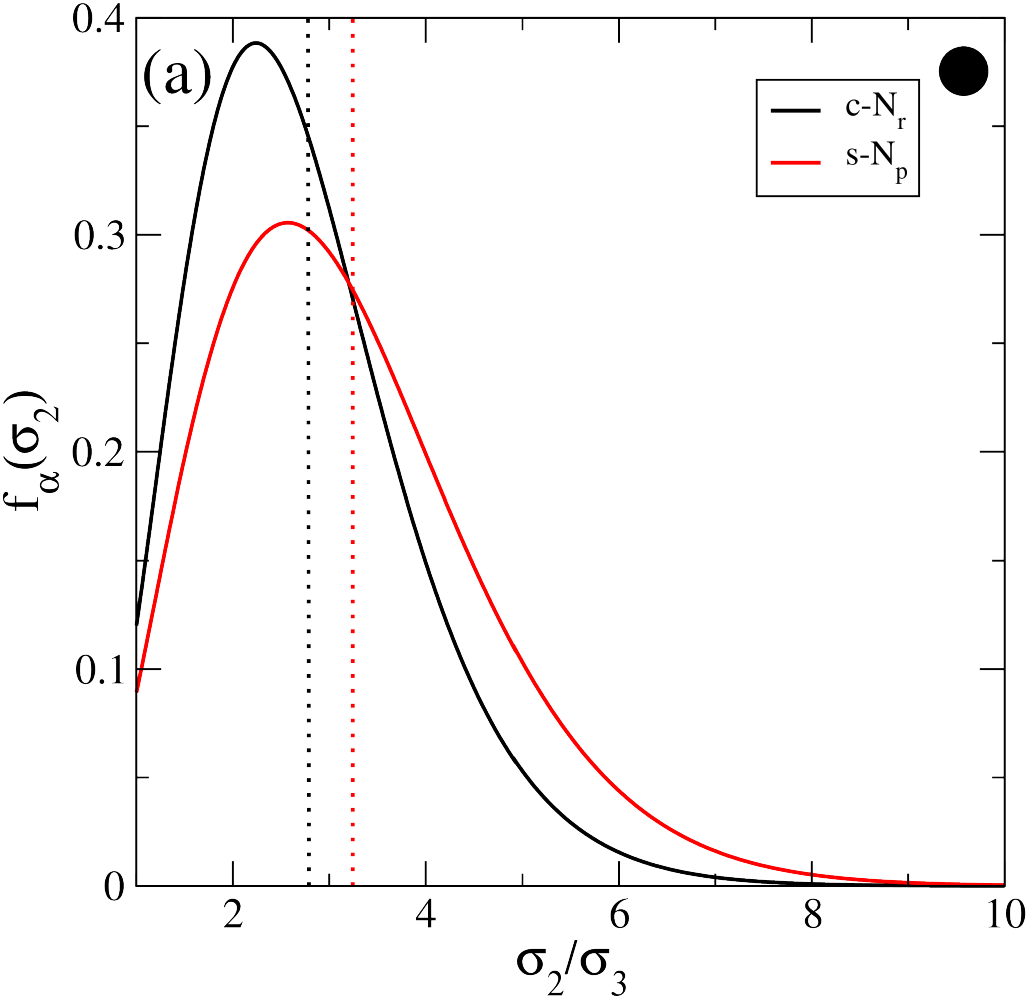}
        \includegraphics[width=2.3in]{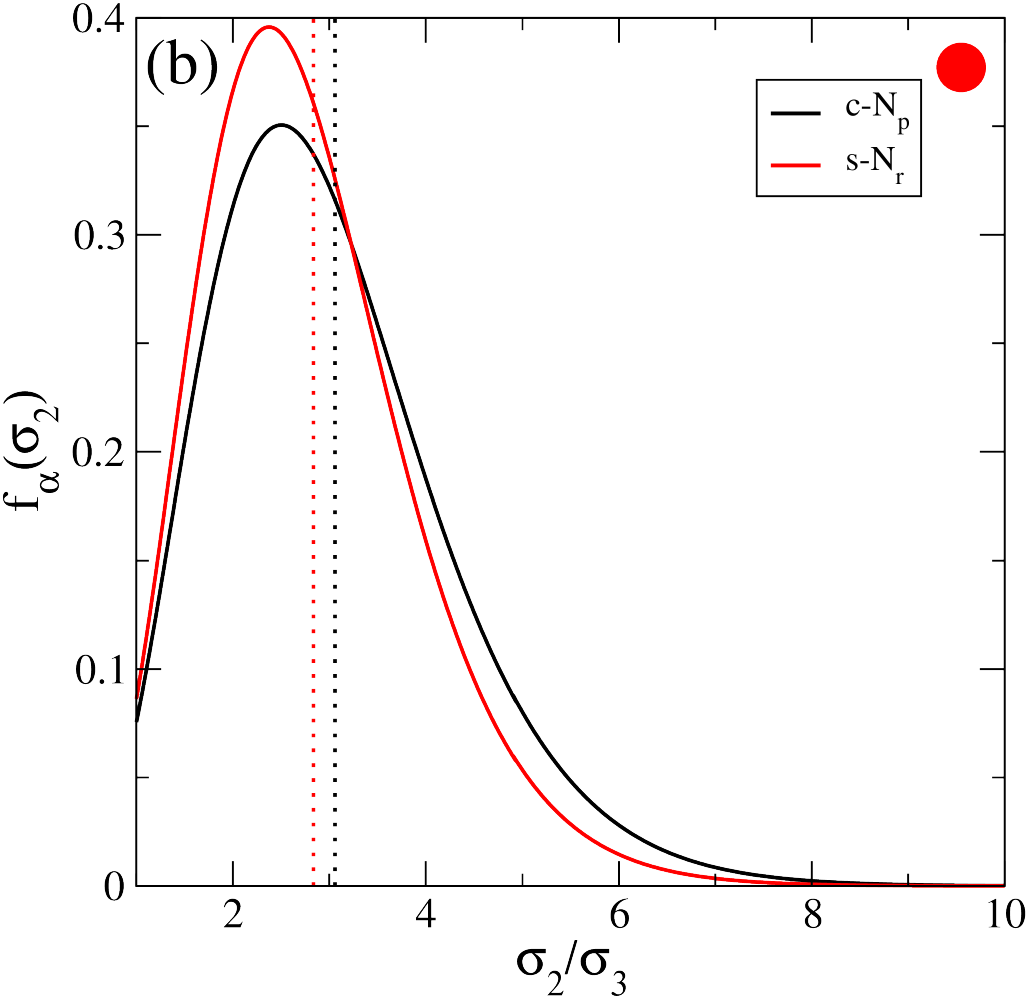}
        \includegraphics[width=2.3in]{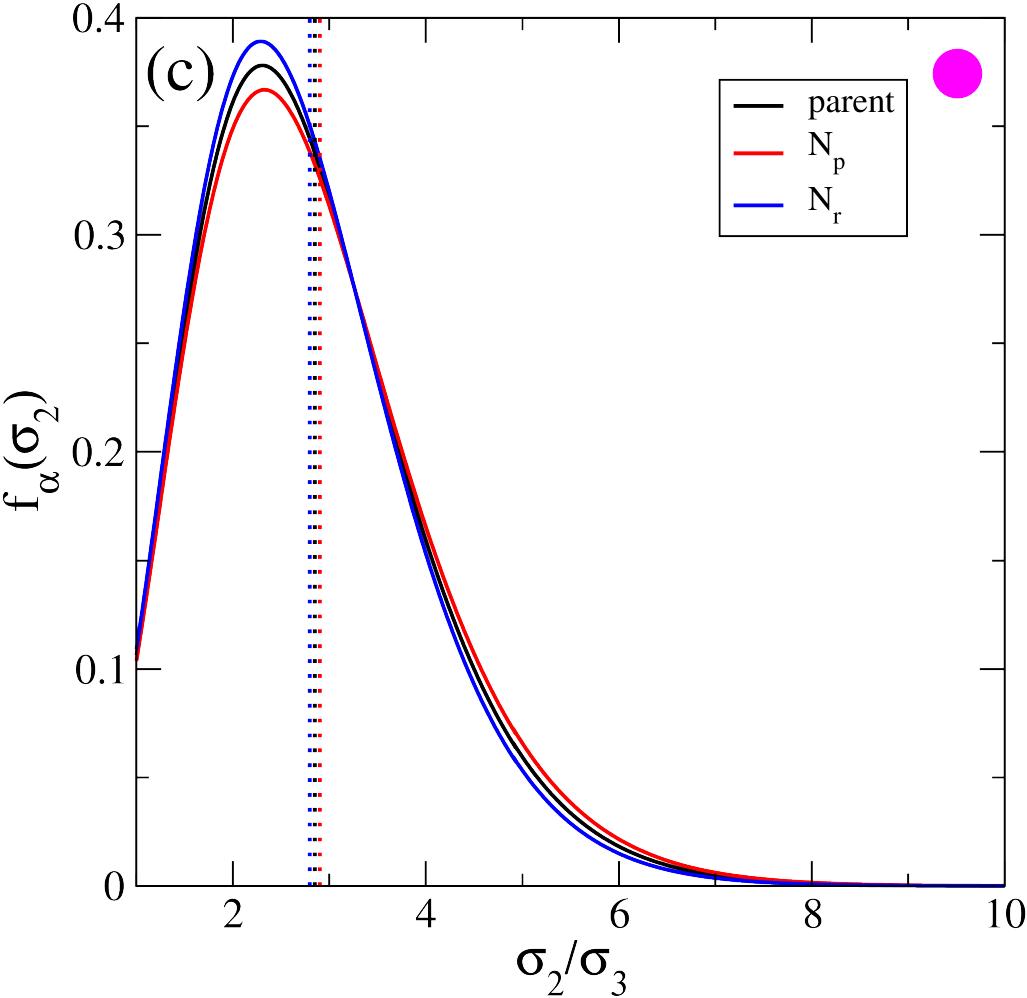}
        \caption{Coexisting intermediate-length distribution functions $f_{\alpha}(\sigma_2)$ for the same states shown in the phase diagram of
        Fig.~\ref{fig6}. Vertical lines indicate the mean values $\langle \sigma_2\rangle_{\alpha}$, illustrating  the fractionation of plate-like and rod-like particles
        between the two coexisting phases: the N$_{\rm p}$ and N$_{\rm r}$ phases
        are enriched in plate-like and rod-like particles, respectively. As the coexistence phases approach the tetracritical point, fractionation becomes progressively weaker.
        }
        \label{fig7}
\end{figure*}

The N$_{\rm r}$-N$_{\rm p}$ demixing transition exhibits even stronger fractionation, as 
can be readily understood from Fig.~\ref{fig6}, where we show a representative demixing transition for the particular case $s=0.4$. Panel (a) shows the quantities characterizing the parent phase, namely the
packing fraction $\eta_0$ and mean intermediate length $\sigma_0$, which are the relevant control parameters from an experimental point of view.
In experiments, the preparation of colloidal samples generally results in polydisperse particle-size distributions that cannot be precisely controlled.  
Their mean and standard deviation are measured after sample preparation. Thus, 
each sample is characterized by a parent distribution with a given mean 
$\sigma_0$ and polydispersity coefficient $s$. The packing fraction $\eta_0$ can then be varied by adding solvent, thereby following the so-called dilution line. Consequently, a given sample follows a vertical  
line in panel (a), corresponding to a fixed value of 
$\sigma_0$. 

In Fig.~\ref{fig6} we have chosen to plot the entire N$_{\rm r}$--N$_{\rm p}$ demixing transition, 
including the portion that becomes metastable at packing fractions above the N$_{\alpha}$--B transition, since our aim is to illustrate the underlying phenomenology. In Fig.~\ref{fig6}(a) each curve corresponds to the values of $(\sigma_0,\eta_0)$ along the N$_{\rm r}$--N$_{\rm p}$ coexistence region for a fixed value of the fraction $t={0,\ 0.15,\ 0.3,\ 0.5,\ 0.7,\ 0.85,\ 1}$ of the total volume occupied by the N$_{\rm p}$ phase.

Let us select a value of $\sigma_0$ slightly to the right of the multicritical point and 
move from low densities within the region of stability of the 
N$_{\rm p}$ phase. The first intersection with the curve $t=1$ indicates the 
value of $\eta_0$ at which the N$_{\rm p}$ phase, occupying essentially the entire sample volume, first coexists  with a vanishingly small amount of the N$_{\rm r}$ phase. Upon further increasing $\eta_0$, the vertical line successively crosses the curves with different values of $t$, until it reaches the $t=0$ curve. At this point 
the N$_{\rm r}$ phase occupies the entire sample and coexists with an infinitesimal amount of the N$_{\rm p}$ phase, thus completing the N$_{\rm p}$--N$_{\rm r}$ transition.
A further increase in $\eta_0$ takes the system into the  
N$_{\rm r}$ stability region, until intersection of the $t=0$ curve for a second time. 
Note how the curve bends backwards towards lower values of $\sigma_0$, marking the onset of another coexistence region in which the roles of N$_{\rm r}$ and N$_{\rm p}$, regarding coexistence, 
are interchanged. However, the vertical dilution line intersects only a few curves corresponding to relatively small values of $t$, indicating that this new transition cannot be completed within the range of uniform-phase coexistence shown. 

Of course, this discussion assumes that the uniform phases remain stable 
with respect to NU phases. At sufficiently high packing fractions NU phases will eventually become stable 
and must therefore be included, together with the B phase, in a complete coexistence calculation. The topology of the curves also shows that, for values of $\sigma_0$ to the right of the turning point [indicated by the square in panel (a)], 
any two-phase coexistence between uniform N phases cannot be completed at moderate densities before the onset of stability of the NU phases. 

In Fig.~\ref{fig6}(b) we show the coexisting binodals corresponding to 
each value of $t$, where the packing fraction and mean intermediate lengths now refer to 
the coexisting phases rather than to the parent phase. Note the central instability 
gap separating the N$_{\rm r}$ (left) and N$_{\rm p}$ (right) branches of the binodals. This separation provides a clear signature of the strong fractionation of 
rod-like and plate-like particles between the N$_{\rm r}$ and N$_{\rm p}$ 
coexisting phases, respectively.   

To illustrate the fractionation effect more clearly, Fig.~\ref{fig7} shows  
the length-distribution functions of the coexisting phases corresponding to the points marked by circles in Fig.~\ref{fig6}, belonging to the $t=0$, $t=1$ 
and $t=0.5$ curves. Note that, for the latter case, the two coexisting distributions differ only slightly from 
the parent distribution and are also very similar to each other, as expected from the proximity 
of this state to the multicritical one. Figure~\ref{fig7} clearly shows that the N$_{\rm r}$-distribution is
shifted towards smaller values of $\sigma_2$ relative to the N$_{\rm p}$ distribution, while the latter exhibits a slower decay at large $\sigma_2$. This reflects the preferential partitioning of more rod-like and more plate-like particles into the N$_{\rm r}$ and N$_{\rm p}$ phases, respectively.
It is interesting to note that, 
despite the similarity between the two coexisting distributions [see panel (c)], their
differences are sufficient to drive a demixing transition. Such similarity between the coexisting distributions is typical of the demixing regions found in the phase diagrams shown in Fig.~\ref{fig3} (b)-(f).

\subsection{Computer simulations}
\label{results_MC}

\begin{figure*}
        \includegraphics[width=0.99\textwidth]{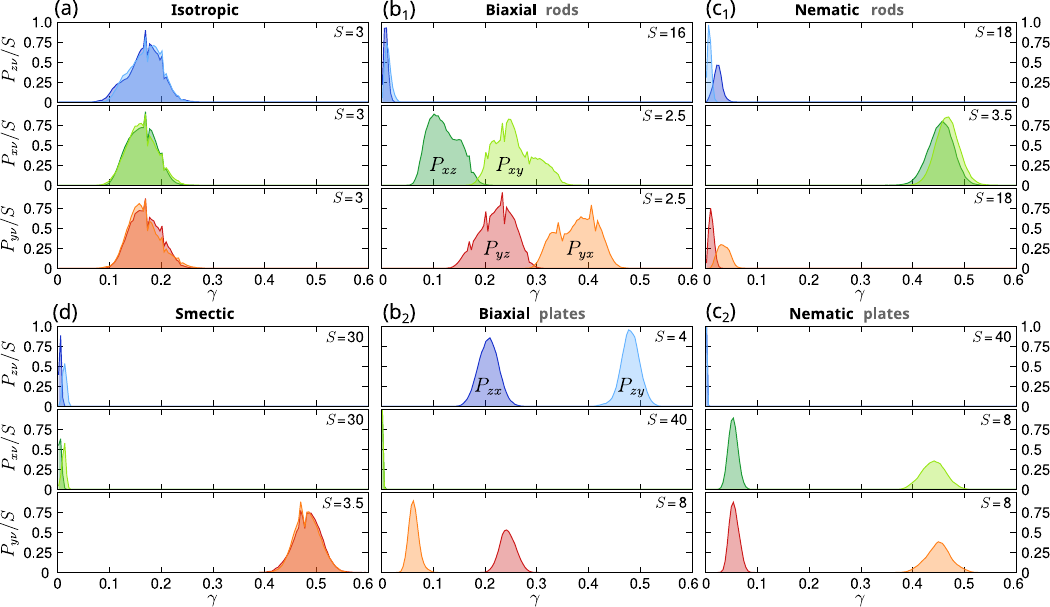}
        \caption{Scaled probability distributions, $P_{\mu\nu}/S$ where $\mu,\nu\in\{x,y,z\}$, of the molar fractions $\gamma_{\mu\nu}$ in different phases obtained from Monte Carlo simulations.
        The scaling factor, $S$, varies across panels and is indicated within each plot.
        The color code of the distribution functions matches the species representation in Fig.~\ref{fig1} and Fig.~\ref{fig2}.
        The phases match those shown in Fig.~\ref{fig2}:
        (a) I fluid at packing fraction $\eta_0=0.23$ and average intermediate length $\sigma_0=2.4$,
        (b$_1$) B phase of rods ($\eta_0=0.32$ and $\sigma_0=2.4$),
        (b$_2$) B phase of plates ($\eta_0=0.42$ and $\sigma_0=2.7$),
        (c$_1$) N$_{\rm r}$ phase ($\eta_0=0.34$ and $\sigma_0=2.15$),
        (c$_2$) N$_{\rm p}$ phase ($\eta_0=0.39$ and $\sigma_0=3.85$),
        and (d) rod-rich Sm phase ($\eta_0=0.5$ and $\sigma_0=2.4$).
        }
        \label{fig8}
\end{figure*}

To validate and complement the theoretical results, we performed canonical Monte Carlo (MC) computer simulations.
The system consists of $N=1000$ particles initialized in a cubic box with periodic boundary conditions.
Given the relatively small system size, finite-size effects cannot be entirely ruled out.
For selected values of $\sigma_0$ we also performed simulations with $N=2000$, and the results did not change significantly.
Particle positions and orientations are initially randomized in a highly diluted system where the simulation box side length corresponds to a packing fraction $\eta_0 = 0.015$.
That is, deep within the isotropic state.
The parent distributions of intermediate particle lengths, $\sigma_2$, in the simulations are discretized versions of those implemented in the theory.
We use a discretization step of $0.25\sigma_1$ for $N=1000$ and reduce it proportionally for $N=2000$.
The required number of particles in each interval is calculated to reproduce the target parent distribution.
The specific value of $\sigma_2$ for each particle is chosen uniformly at random within each interval.
We simulate systems with a polydisperse coefficient $s=0.49$.

Three types of MC moves are executed during the simulations: (i) simultaneous translation and rotation of a particle, (ii) swaps between two randomly chosen particles, and (iii) isotropic volume compressions of the simulation box.

Simulations begin by compressing the box from the initial dilute system to an isotropic state with $\eta_0=0.22$.
In each compression attempt, the side length of the box is scaled by a factor of $1-\Theta_{\rm c}$ where $\Theta_{\rm c}$ is a random variable uniformly distributed between $0$ and $0.005$.
Between consecutive compression attempts, we perform $10^2$ Monte Carlo steps (MCS) and $10^1N$ particle swap attempts.
One MCS is an attempt to translate and rotate $N$ randomly chosen particles in the system.
For rotations, a new orientation is chosen at random.
Hence, the particle orientation remains unchanged one out of six times.
Once the desired packing fraction is reached, the system is evolved for $10^6$ MCS and $10^5N$ swap attempts to achieve equilibration, sample the distribution functions, and compute the orientational order parameters.

Following this protocol (first compression next data production), the packing fraction is increased at intervals of $\Delta\eta_0=0.01$.
Finally, to mitigate the effect of statistical fluctuations, the orientational order parameters at each packing fraction are averaged over ten independent simulation runs.
Note that, due to the discrete nature of the parent distribution, the value of $\sigma_0$ fluctuates slightly between runs with different initial configurations, despite sharing the same target parent distribution of intermediate lengths.

Characteristic snapshots of resulting particle configurations in different phases are presented in Fig.~\ref{fig2}, with their corresponding orientational probability distributions, $P_{\mu\nu}(\gamma)$, shown in Fig.~\ref{fig8}.
For clarity, we plot scaled distributions $P_{\mu\nu}/S$, where the individual scaling factors $S$ are specified in each panel.
Here, $P_{\mu\nu}(\gamma)$ is the probability of finding species $\mu\nu$ with molar fraction $\gamma$ during the simulation.
In the isotropic state, Fig.~\ref{fig8}(a), all probability distributions are centered around $\gamma=\frac16$.
For both uniaxial nematics, Figs.~\ref{fig8}(c$_1$) and~\ref{fig8}(c$_2$), as well as for the smectic of rods, Fig.~\ref{fig8}(d), the six species group into three distinct distributions types as dictated by symmetry.
For example, in the uniaxial nematic of plates shown in Fig.~\ref{fig8}(c$_2$), we find $P_{xy}\approx P_{yx}$, $P_{xz}\approx P_{yz}$, and $P_{zx}\approx P_{zy}$.
In contrast, all six distributions are distinct in the biaxial phases, see Figs.~\ref{fig8}(b$_1$) and~\ref{fig8}(b$_2$).

Most distribution exhibit a Gaussian-like shape, providing strong evidence that the system has reached an equilibrium state.
An exception is the biaxial of rods, Fig.~\ref{fig8}(b$_1$).
In some simulations, we observe bimodal rather than unimodal distributions, see e.g.~$P_{xy}$ in Fig.~\ref{fig8}(b$_1$).
This behaviour might be attributed to a global reorientation of one of the directors during the course of the simulation.
Alternatively, bimodal distributions can signal the onset of clustering, fractionation, or demixing between two coexisting phases in the system.
Our small scale simulations were only intended to confirm the global features of the theoretical phase diagrams.
The reduced system size precludes a definitive conclusion regarding the presence of phase separation in the system.

\begin{figure*}
    \centering
    \includegraphics[width=0.99\textwidth]{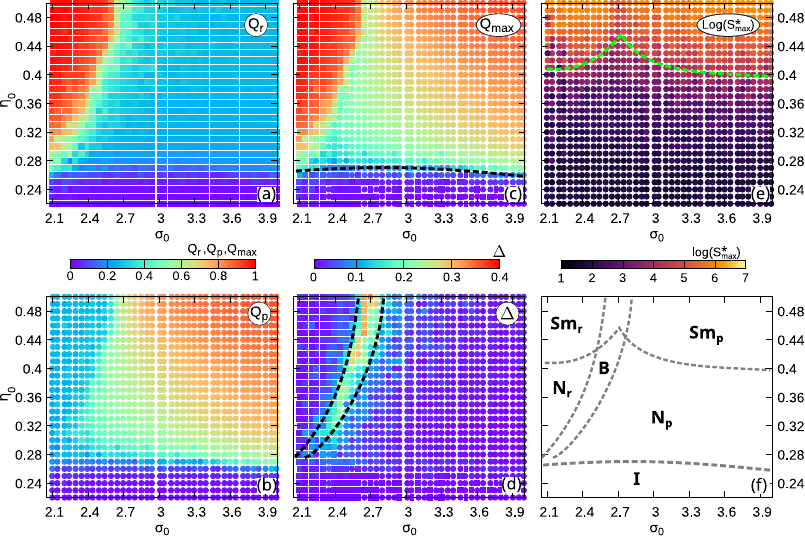}
    \caption{
        Order parameters in the plane of average intermediate length, $\sigma_0$, and global packing fraction, $\eta_0$, according to Monte Carlo simulations ($N=1000$).
        The polydisperse coefficient is $s=0.49$.
        (a) Uniaxial order parameter of rods, $Q_{\rm r}$.
        (b) Uniaxial order parameter of plates, $Q_{\rm p}$.
        (c) Maximum uniaxial order parameter $Q_{\rm max}=\max(Q_{\rm r},Q_{\rm p})$.
        (d) Biaxial order parameter, $\Delta$.
        (e) Logarithm of the maximum value of the scaled structure factor, $S_{\max}^* = S_{\rm max} / \sigma_3$.
        In panels (a) to (d), each symbol represents an average over $10$ independent simulations.
        In panels (c) and (d), square symbols denote regions where the uniaxial order parameter of rods dominates ($Q_{\rm r}>Q_{\rm p}$), whereas circles indicate regions where the plates dominate  ($Q_{\rm p}>Q_{\rm r}$).
        (f) State diagram showing the approximated stability regions of several phases: I, N$_{\rm r}$, N$_{\rm p}$, Sm of rods (Sm$_{\rm r}$) and Sm of plates (Sm$_{\rm p}$).
    }
    \label{fig9}
\end{figure*}

We sampled several order parameters to estimate the phase boundaries from the simulation data.
Figures~\ref{fig9}(a) and~\ref{fig9}(b) show the uniaxial order parameter for rods $Q_{\rm r}$ and plates $Q_{\rm p}$, respectively, in the plane of average intermediate length $\sigma_0$ and total packing fraction $\eta_0$.
Each symbol in these plots represents the order parameter averaged over ten independent runs.
As expected, both $Q_{\rm r}$ and $Q_{\rm p}$ increase by increasing the global packing fraction.
Decreasing $\sigma_0$ renders the particles more rod-like, and hence $Q_{\rm r}$ begins to grow at lower packing fractions.
Analogously, the rapid increase of $Q_{\rm p}$ shifts to lower packing fractions for more plate-like particles, i.e. for larger values of $\sigma_0$.

In Fig.~\ref{fig9}(c), we plot the maximum of the two uniaxial order parameters, $Q_{\rm max}=\max(Q_{\rm r},Q_{\rm p})$.
We use $Q_{\rm max}$ to delineate the approximate boundary between isotropic and nematic states as the region where the uniaxial order parameter increases rapidly, indicated by the black dashed lines in Fig.~\ref{fig9}(c).
To distinguish between the rod-like and plate-like nematic phases in Fig.~\ref{fig9}(c), we use square symbols when $Q_{\rm r}>Q_{\rm p}$ and circles when $Q_{\rm p}>Q_{\rm r}$.

Next, to identify the approximate boundaries of the biaxial phase, we plot the biaxial order parameter $\Delta$ in Fig.~\ref{fig9}(d).
We present either $\Delta_{\rm r}$ (squares) if $Q_{\rm r}>Q_{\rm p}$ or $\Delta_{\rm p}$ (circles) if $Q_{\rm p}>Q_{\rm r}$.
The results reveal a well defined region where $\Delta$ is significantly greater than the background level of the isotropic state, even though it does not reach values comparable to those of the uniaxial order parameter (see the color bars).

To identify the onset of non-uniform spatial phases, we calculated the structure factor of the particle configurations.
Figure~\ref{fig9}(e), shows the maximum value of the structure factor (scaled with $\sigma_3$) on a logarithmic scale.
As before, each symbol represents an average over ten independent realizations.
For each realization, the final microstate was used to compute the structure factor as a function of the reciprocal-space wave vector, and to extract its maximum value (excluding the zero-wave vector mode).
A rapid increase of the structure factor is observed near $\eta_0\sim 0.4$.
Furthermore, this growth shifts to slightly higher, yet noticeable, packing fractions within the stability region of the biaxial phases.
The characteristic wavelength $d$ corresponding to the maximum value of the structure factor (not shown) is fully compatible with a smectic of rods ($d\approx\sigma_1$) in the region where $Q_{\rm r}$ dominates, and with a smectic of plates $d\approx\sigma_3$ where $Q_{\rm p}$ dominates.

Finally, Fig.~\ref{fig9}(f) presents the state diagram constructed from our Monte Carlo simulations using the phase boundaries discussed above.
We reiterate that because these boundaries are approximate, fine details cannot be definitively resolved.
This includes the exact topology of the region where isotropic, uniaxial, and biaxial phases emerge, as well as the occurrence of demixing.
Similarly, while the existence of both plate-like and rod-like biaxial states is clear, we cannot differentiate their individual stability regimes within the biaxial region.
Nevertheless, the global features of the simulated state diagram are robust and qualitatively consistent with the theoretical predictions, cf. Fig.~\ref{fig3}(f) and Fig.~\ref{fig9}(f).
The phase boundaries according to MC simulations are shifted toward higher packing fractions relative to the theoretical predictions.
This shift is expected because our theoretical approach relies on scaled particle theory, which is known to overestimate exclude volume effects.

\section{Conclusions}
\label{conclusions}

We have investigated the phase behaviour of continuously polydisperse mixtures of hard board-like particles within the restricted-orientation (Zwanzig) model, in which the particle orientations are limited to the six possible permutations of the principal particle axes. Continuous polydispersity was introduced in the intermediate edge length of the particles, and the phase behaviour was studied using a density-functional theory based on FMT. In addition, we performed MC simulations in the canonical (NVT) ensemble to assess the qualitative accuracy of the theoretical predictions.

One of the main motivations of this work was to incorporate the calculation of the N-N demixing transition into the theoretical description. A second objective was to investigate how continuous polydispersity in the intermediate edge length of the particles affects the stability of the biaxial nematic phase.
We find that the N$_{\rm r}$-N$_{\rm p}$ demixing transition indeed emerges above the multicritical point, from which the isotropic-nematic coexistence lines, I-N$_{\rm r}$ and I-N$_{\rm p}$, also originate.
Somewhat unexpectedly, the demixing region always occupies a narrow portion of the phase diagram, even for small polydispersities of about 10$\%$. Although this region broadens with increasing polydispersity, its growth is significantly smaller than that of the B-phase stability region, which is bounded from below by the B-N$_{\rm p}$ and from above by the B-Sm transition.
We therefore conclude, in agreement with the findings of Ref. \cite{Patti1}, that continuous size polydispersity enhances the stability of the B phase. 

When the polydispersity 
increases, the multicritical point moves to 
lower values of the mean intermediate length $\sigma_0$ and packing fraction $\eta_0$
as a consequence of the amplified entropic excluded volume effect driven by the increased 
presence of plate-like particles. The Sm phases of rods or plates limit from above
the region of stability of the N$_{\rm r}$ and N$_{\rm p}$ phases, 
respectively. These Sm phases have a bifurcation, 
for any $\sigma_0$, with a relative constant period, compatible with the largest and smallest edge-lengths respectively. We conclude that the Sm phases accept high values of intermediate length polydipersities. 

The preceding results concern phase diagrams calculated for five different values of the polydispersity coefficient $s$, with the largest edge length of the boards fixed at $\sigma_1=10$. We have also calculated a phase diagram for $\sigma_1=5$ and $s=0.42$. The main result is that, at moderate packing fractions, the B phase loses stability with respect to N$_{\rm r}$--N$_{\rm p}$ demixing, although most of the corresponding demixing region is metastable with respect to transitions to nonuniform phases. At high packing fractions, the coexistence path involves, in addition to N$_{\rm r}$--N$_{\rm p}$ demixing, B--N$_{\rm p}$ and B--N$_{\rm r}$ transitions. These coexistence paths can only be completed when $\sigma_0$ lies within a certain range. Their topology is therefore strongly influenced by the specific form of the parent distribution function.

Our Monte Carlo simulations, carried out for the values $\sigma_1=10$ and $s=0.49$, qualitatively validate our theoretical results.
The topology of the simulated phase diagram agrees with the theoretical predictions.
The simulations confirmed the existence of a region of stability of the biaxial phase at intermediate values of the global packing fraction and mean intermediate length.
We also found I-N$_{\rm r}$ and I-N$_{\rm p}$ transitions by increasing the packing fraction, as well as N$_{\rm r}$-B and N$_{\rm p}$-B transitions by changing the mean intermediate length at constant packing fraction. 
The uniaxial and biaxial nematic phases are replaced at high packing fractions by two types of smectic phases.
For low $\sigma_0$, the smectic layers are formed by rod-like particles with major axes perpendicular to the layers, so the smectic period is close to $\sigma_1$. 
For high values of $\sigma_0$, the layers are formed by plate-like particles and hence the smectic period is compatible with the smallest edge length of the boards, $\sigma_3$.
The phase diagram obtained from MC simulations is shifted toward higher packing fractions when compared to that obtained from the theory.
This quantitative difference is due to the fluid pressure overestimation given by the Scaled Particle Theory (the uniform bulk limit of our density functional theory).
The simulated phase diagram is not precise enough to resolve the topology in the region where the biaxial phase emerges.
Finally, we have found some clustering and the occurrence of quasi-bimodal probability distributions of molar fractions, which are consistent with the existence of N$_{\rm r}$-N$_{\rm p}$ demixing. 
However, due to the reduced size of the simulated systems, we cannot confirm the existence of demixing.

Monte Carlo simulations of the one-component fluid of freely rotating hard-board particles showed an N$_{\rm p}$-B transition for large values of the largest aspect ratio, $\kappa_1=\frac{\sigma_1}{\sigma_3}> 23$, with the secondary aspect ratio, $\kappa_2=\frac{\sigma_2}{\sigma_3}$, fixed close to the dual shape value, $\displaystyle{\kappa_2=\sqrt{\kappa_1}}$ \cite{Roij1}. 
Similar behaviour was reported for rhombic platelets and triangular prisms, for which 
the B-phase stability region lies well above the I phase for all values of the relevant
geometric parameters \cite{Roij1}. Recently the FMT formalism was applied to study 
biaxial ordering in freely rotating spherotriangles \cite{Wittmann}. The phase diagram exhibits a B-phase region of stability above the uniaxial N phase, and well separated from the I-phase stability region \cite{Wittmann}. By contrast,
the phase diagrams of the polydisperse Zwanzig model obtained from FMT (Fig. \ref{fig3}) and from MC simulations (Fig. \ref{fig9}) show that, for sufficiently large values of $\kappa_1$, although still considerably smaller than 23, a direct transition from the I fluid to the B phase becomes possible. According to the theory, this direct transition I--B transition occurs at the value of $\kappa_2$ corresponding to the multicritical point, whereas MC simulations indicate a rather narrow interval of $\kappa_2$ over which a direct I-B transition takes place. Its precise location, however, remains uncertain because of finite-size effects. We can therefore conclude that the restriction in particle orientations inherent in the Zwanzig model is responsible for this qualitative difference in the topology of the phase diagram.

For future research, we plan to extend the present model to investigate the effects of polydispersity on sedimentation behavior. 
The external gravitational field can be incorporated into the theoretical framework either by generalizing sedimentation path theory~\cite{Eckert1,Eckert2} to size-polydisperse systems or by directly minimizing an inhomogeneous density functional~\cite{Eckert3}.
In presence of gravity, the particle distribution profile becomes non-uniform along the vertical coordinate of the sedimentation cuvette.
Hence, an interesting open question is whether the gravitational field can stabilize a biaxial phase within the cuvette, even under conditions where such a phase remains unstable in the bulk.

\appendix 

\section{Spinodal instabilities}
\label{app1}

Here we implement a bifurcation analysis of the equation obtained by density 
functional minimization of the free-energy with respect to the density profiles 
$\rho_{\mu\nu}(\sigma_2,{\bm r})$, which now depend on the spatial variable ${\bm r}$. This equation,
\begin{eqnarray}
	\rho_{\mu\nu}(\sigma_2,{\bm r})=\rho_{\mu\nu}(\sigma_2) e^{\Delta c_{\mu\nu}(\sigma_2,{\bm r})},
	\label{departure}
\end{eqnarray}
relates the density profiles and the difference, $\Delta c_{\mu\nu}(\sigma_2,{\bm r})$, between the one-body 
direct correlation functions of the nonuniform, $c_{\mu\nu}(\sigma_2,{\bm r})$, and uniform, 
$c_{\mu\nu}(\sigma_2)$, phases, and allows for the calculation of the bifurcation condition.

We assume that, close to the bifurcation, the following first-order approximation to the density profile 
expansion with respect to the small functions $\epsilon_{\mu\nu}(\sigma_2,{\bm r})$ can be applied:
\begin{eqnarray}
	\rho_{\mu\nu}(\sigma_2,{\bm r})=\rho_{\mu\nu}(\sigma_2)\left[1+\epsilon_{\mu\nu}(\sigma_2,{\bm r})
	\right]. \label{insert}
\end{eqnarray}
Inserting (\ref{insert}) into  Eqn. (\ref{departure}), taking into account that  
\begin{eqnarray}
	-c_{\mu\nu}(\sigma_2,{\bm r})=\sum_{\alpha} 
	\int d{\bm r}' \frac{\partial \Phi_{\rm exc}}{\partial n_{\alpha}}({\bm r}')\omega^{(\alpha)}_{\mu\nu}
	(\sigma_2,{\bm r}-{\bm r}'),
\end{eqnarray}
where the weighted densities are given by
\begin{eqnarray}
	n_{\alpha}({\bm r})=\sum_{\mu,\nu}\int_{\sigma_3}^{\sigma_1} d\sigma_2 \int d{\bm r}'\rho_{\mu\nu}(\sigma_2,{\bm r}') \omega^{(\alpha)}_{\mu\nu}
	(\sigma_2,{\bm r}-{\bm r}'), \nonumber\\
\end{eqnarray}
and the excess part of the 
free-energy density $\Phi_{\rm ex}({\bm r})$, given by (\ref{excess}), now depends on ${\bm r}$, it is easy to obtain the equation 
\begin{eqnarray}
	&&-\Delta c_{\mu\nu}(\sigma_2,{\bm r})= 
	\sum_{\alpha,\beta} \Phi_{\alpha\beta}\sum_{\tau, \theta}
	\int_{\sigma_3}^{\sigma_1}d\sigma_2'\int d{\bm r}'\int d{\bm r}'' \nonumber\\
	&&\times \omega_{\mu\nu}^{(\alpha)}(\sigma_2,{\bm r}-{\bm r}') 
	\omega_{\tau\theta}^{(\beta)}(\sigma_2',{\bm r}'-{\bm r}'') 
	\rho_{\tau\theta}(\sigma_2')\epsilon_{\tau\theta}(\sigma_2',{\bm r}''),\nonumber\\ 
	\label{chorizo}
\end{eqnarray}
valid up to first order. Here we have used the shorthand notation $\displaystyle{\Phi_{\alpha\beta}=\frac{\partial^2 \Phi_{\rm exc}}{
	\partial n_{\alpha}\partial n_{\beta}}}$ for the 
    second derivative of $\Phi_{\rm exc}$ with respect to the 
    weighted densities, evaluated at the uniform 
    phase condition.

Inserting (\ref{chorizo}) into (\ref{departure}), and taking Fourier transform in the 
Taylor-expansion of the exponential up to first order, we  obtain 
\begin{eqnarray}
	&&\hat{\epsilon}_{\mu\nu}(\sigma_2,{\bm q})=-\sum_{\alpha,\beta}\Phi_{\alpha\beta} 
	\hat{\omega}^{(\alpha)}_{\mu\nu}(\sigma_2,{\bm q}) u_{\beta}({\bm q}), \label{epsilon}\\
	&&u_{\beta}({\bm q})\equiv \sum_{\tau,\theta} \int_{\sigma_3}^{\sigma_1}d\sigma_2'
	\rho_{\tau\theta}(\sigma_2') \hat{\omega}^{(\beta)}_{\tau\theta}(\sigma_2',{\bm q}) 
	\hat{\epsilon}_{\tau\theta}(\sigma_2',{\bm q}), \label{definition}\nonumber\\
\end{eqnarray}
where the ``hat'' over all quantities indicates a Fourier transform of the original ones. In these equations ${\bm q}$ is the 
wave vector.

Multiplying Eqn.~(\ref{epsilon}) by $\rho_{\mu\nu}(\sigma_2)\hat{\omega}^{(\gamma)}_{\mu\nu}(\sigma_2,{\bm q})$, 
integrating over $\sigma_2$, summing over species $\mu\nu$, and using the definition (\ref{definition}), we 
obtain 
\begin{eqnarray}
	&&u_{\gamma}({\bm q})=-\sum_{\alpha,\beta} {\cal N}_{\gamma\alpha}({\bm q}) \Phi_{\alpha\beta} u_{\beta}({\bm q}), 
	\label{las_u}\\
	&&{\cal N}_{\gamma\alpha}({\bm q})\equiv \sum_{\mu,\nu}\int_{\sigma_3}^{\sigma_1} d\sigma_2 \rho_{\mu\nu}
	(\sigma_2) \hat{\omega}^{(\gamma)}_{\mu\nu}(\sigma_2,{\bm q})\hat{\omega}_{\mu\nu}^{(\alpha)}
	(\sigma_2,{\bm q}).\nonumber\\
\end{eqnarray}
Equation~(\ref{las_u}) can be written in matrix form as 
\begin{eqnarray}
	H({\bm q}) {\bm u}({\bm q})=\left(I+{\cal N}({\bm q})\cdot \Phi\right){\bm u}({\bm q})=0,
	\label{matrix}
\end{eqnarray}
with $I$ the identity matrix. The $8\times 8$ matrices ${\cal N}({\bm q})$ and $\Phi$ have elements ${\cal N}_{\gamma\alpha}({\bm q})$ and $\Phi_{\alpha\beta}$, respectively. Also, ${\bm u}({\bm q})$ is the column vector of coordinates 
$u_{\beta}({\bm q})$. Equation~(\ref{matrix}) has a nontrivial solution only if 
\begin{eqnarray}
	{\cal H}(\rho_0,{\bm q})\equiv \text{det}\left[H({\bm q})\right]=0.
	\label{bifurca}
\end{eqnarray}
This equation allows us to find the first value of $\rho_0^*$ at which bifurcation between a uniform phase, with equilibrium densities $\rho_{\mu\nu}(\sigma_2)$, and a nonuniform phase takes place. The latter is associated with a wave number ${\bm q}^*$ which results from the 
absolute minimum of ${\cal H}(\rho_0,{\bm q})$ with respect to ${\bm q}$, with the necessary condition
\begin{eqnarray}
	\boldsymbol{\nabla}_{\bm q} {\cal H}(\rho_0,{\bm q})=0. \label{necessary}
\end{eqnarray}
Now selecting ${\bm q}={\bf 0}$ and solving Eqn.~(\ref{bifurca}), the parent number density 
$\rho_0^*$ at which a biaxial nematic phase bifurcates from the uniaxial nematic phase can be obtained.

For completeness, explicit expressions for the Fourier transforms of the 
weighting functions are given:
\begin{eqnarray}
        &&\hat{\omega}_{\mu\nu}^{(0)}(\sigma_2,{\bm q})=\prod_{\tau} 
	\cos\left(\frac{q_{\tau}\sigma_{\mu\nu}^{\tau}}{2}\right),\\
	&&\hat{\omega}_{\mu\nu}^{(3)}(\sigma_2,{\bm q})=\prod_{\tau} \frac{2}{q_{\tau}}
        \sin\left(\frac{q_{\tau}\sigma_{\mu\nu}^{\tau}}{2}\right),\\
	&&\hat{\omega}_{\mu\nu}^{(1\theta)}(\sigma_2,{\bm q})=\frac{2}{q_{\theta}}\sin\left(\frac{q_{\theta}\sigma_{\mu\nu}^{\theta}}{2}\right)
        \prod_{\tau\neq \theta} \cos\left(\frac{q_{\tau}\sigma_{\mu\nu}^{\tau}}{2}\right),\nonumber\\
	&&\\
	&&\hat{\omega}_{\mu\nu}^{(2\theta)}(\sigma_2,{\bm q})=\cos\left(\frac{q_{\theta}\sigma_{\mu\nu}^{\theta}}{2}\right)
        \prod_{\tau\neq \theta} \frac{2}{q_{\tau}}\sin\left(\frac{q_{\tau}\sigma_{\mu\nu}^{\tau}}{2}\right).\nonumber\\
\end{eqnarray}
We numerically solved Eqns. (\ref{bifurca}) and (\ref{necessary}) to find the bifurcation 
values $\rho_0^*$ and ${\bm q}^*$ for the instability of the uniaxial nematic or biaxial nematic phases of rods or plates with respect to inhomogeneities along the principal nematic axes, corresponding to smectic phases of rods or plates. 
In the case where inhomogeneities are parallel to $z$ axis the 
wave number is simply ${\bm q}=(0,0,q)$. To calculate the bifurcation of the biaxial nematic phase 
from the uniaxial nematic phases of 
rods or plates we numerically implemented Eqn.~(\ref{bifurca}), setting ${\bm q}={\bf 0}$. Note that, due to the symmetry ${\cal N}({\bm q})={\cal N}^T({\bm q})$, 36 one-dimensional 
integrals are needed to calculate the elements ${\cal N}_{\gamma\alpha}({\bm q})$ at each step in the evaluation 
of the function ${\cal H}(\rho_0,{\bm q})$. This function in turn is calculated at the equilibrium number density distributions  
$\rho_{\mu\nu}(\sigma_2)$. Therefore, the whole set of equations for the moments, Eqns.~(\ref{set}), has to be solved at each step of the algorithm to find the bifurcation values 
$\rho_0^*$ and ${\bm q}^*$.

\section{Phase diagram for $\sigma_1=5$}
\label{app2}

\begin{figure*}
        \includegraphics[width=3.in]{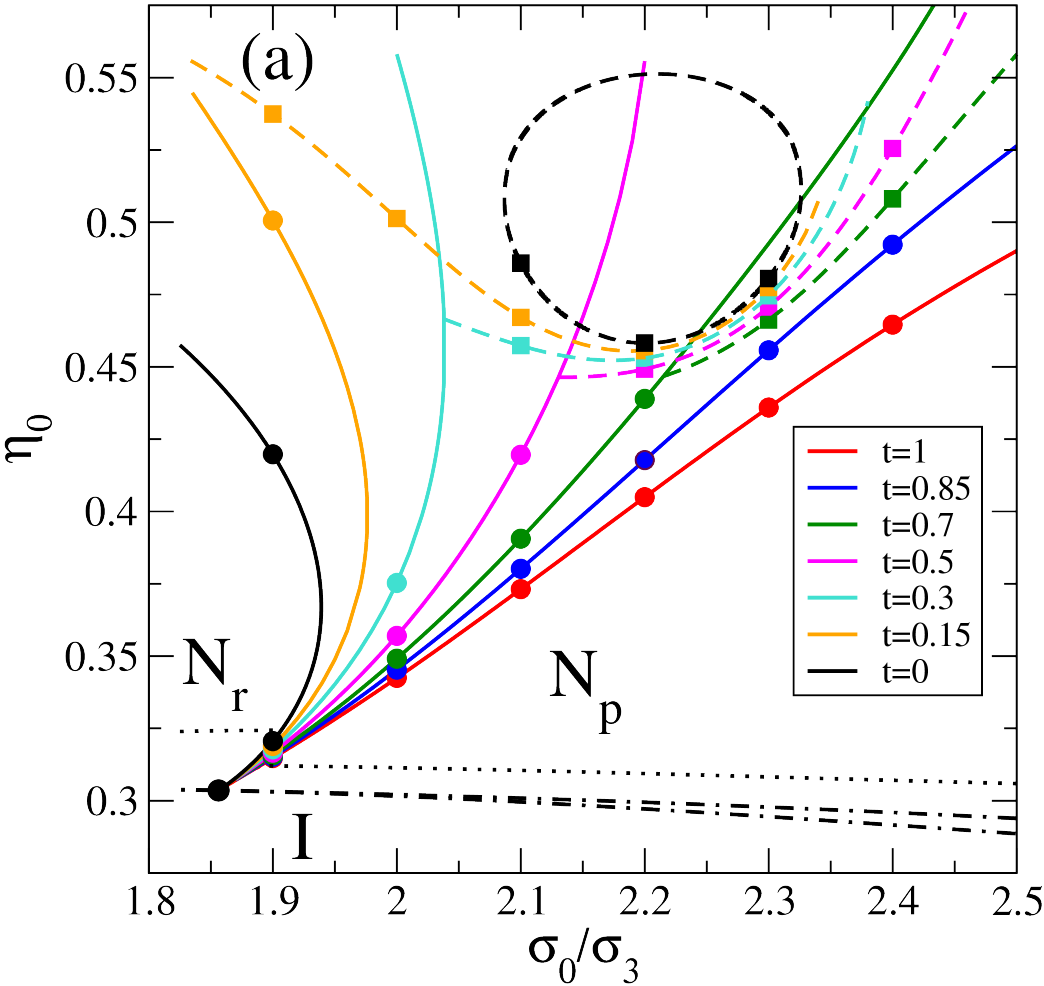}
        \includegraphics[width=3.in]{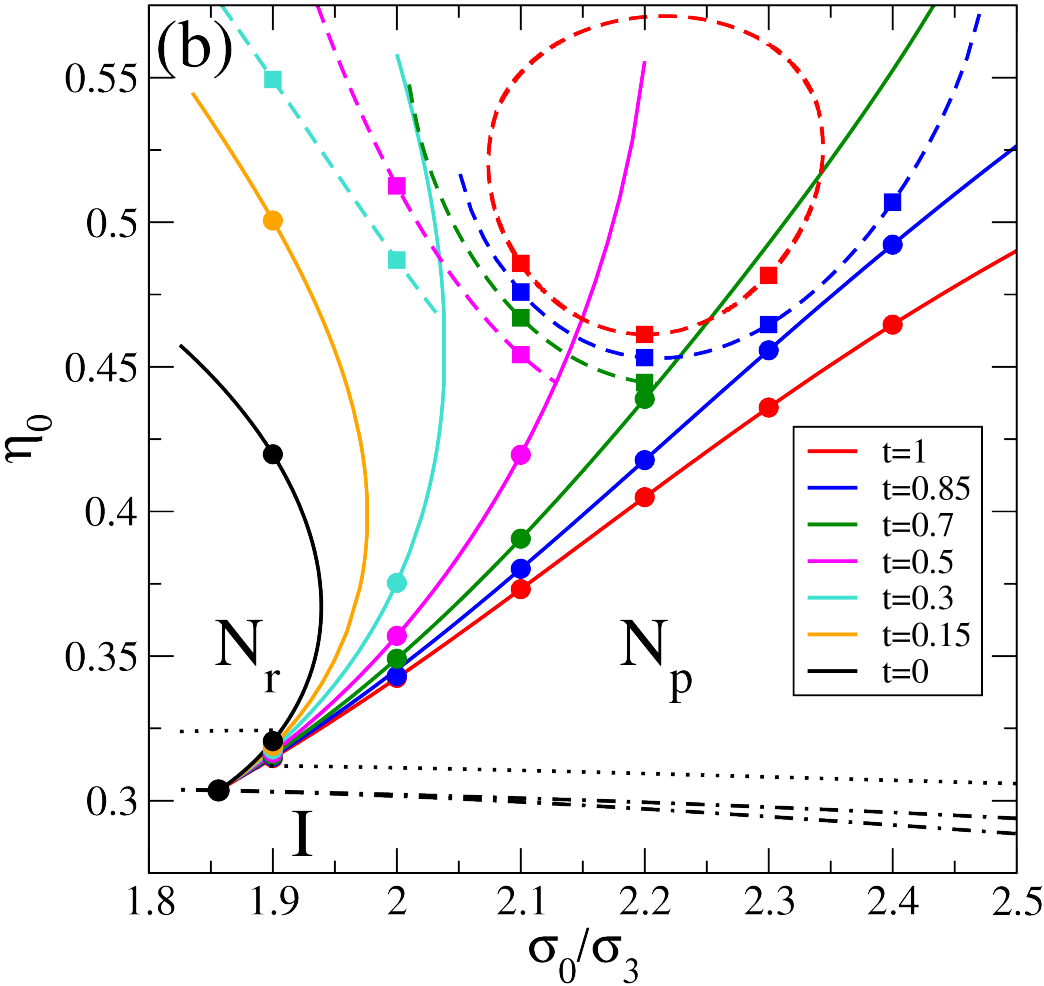}
        \caption{Phase diagram for $\sigma_1=5$ and a fixed polydispersity coefficient $s=0.42$. The curves $\eta_0$ vs. $\sigma_0$ corresponding to N$_{\rm r}$-N$_{\rm p}$ demixing are shown for different values of the volume fraction $t$ occupied by the N$_{\rm p}$ phase. In addition to the I-N transitions (dot-dashed lines) and N$_{\rm r}$-N$_{\rm p}$ demixing (solid lines), the (a) B-N$_{\rm p}$ and (b) B-N$_{\rm r}$ first-order transitions are shown. Dotted lines indicate the N-Sm spinodals.}
        \label{fig10}
\end{figure*}

We have calculated the phase diagram of continuously polydisperse hard board-like particles with the largest edge length fixed at $\sigma_1=5$ and a relatively high degree of polydispersity ($s=0.42$). The resulting phase diagram is shown in Fig. \ref{fig10}, where the parent packing fraction, $\eta_0$, is plotted as a function of the average intermediate edge length, $\sigma_0$. Under these conditions, the B phase becomes unstable with respect to N$_{\rm r}$-N$_{\rm p}$ demixing. This demixing region is, however, confined to a narrow window bounded from below by the multicritical point and from above by the continuous N$_{\rm r,p}$-Sm transition.

We now discuss the demixing transitions that extend into the stability region of NU phases, despite being metastable there. For clarity, the phase diagram is reproduced in two panels, (a) and (b). The main difference between them is that panel (a) shows the B-N$_{\rm p}$ coexistences, whereas panel (b) shows 
the B-N$_{\rm r}$ coexistences, as discussed below. 

Let us first fix $\sigma_0=2.2$ [panel (a)] and increase $\eta_0$, starting from the region of stability of the N$_{\rm p}$ phase. At the value of $\eta_0$ indicated by the red circle on the $t=1$ curve, a first-order phase transition occurs between a cloud N$_{\rm p}$ phase which occupies the entire sample volume, and an incipient shadow N$_{\rm r}$ phase, which occupies a vanishing small fraction of it. As $\eta_0$ is further increased, the coexistence path crosses the $t=0.85$ (blue circle) and $t=0.7$ (green circle) curves, indicating that the volume fraction occupied by the coexisting N$_{\rm r}$ phase increases at the expense of the $N_{\rm p}$ phase. 

However, at a certain point along this path, the N$_{\rm r}$ phase undergoes a 
second-order transition to the B-phase, and the system subsequently follows a 
B-N$_{\rm p}$ path. The magenta square on the dashed curve of the same color corresponds to 
$t=0.5$, where the B-N$_{\rm p}$ phases coexist, each phase occupying half of the total volume). On further increasing $\eta_0$ the path successively crosses the dashed curves corresponding to $t=0.3$ and $t=0.15$, finally reaching the $t=0$ curve at the 
black square. At this point the B phase occupies the entire sample volume and coexists with an infinitesimal amount  of the N$_{\rm p}$ phase, marking the completion of the two-phase coexistence region. 

The same behavior is found 
for $\sigma_0=2.1$ and $\sigma_0=2.3$. However, for values such as 
$\sigma_0=2$ or $\sigma_0=2.4$, the two-phase coexistence 
between uniform phases, either N or B, cannot be completed. This suggests that the NU phases, which have been excluded from the present coexistence calculations but are thermodynamically more stable in this region, may instead coexist with the uniform phases. 

Panel (b) illustrates an alternative scenario, in which
the B-N$_{\rm r}$ coexistences curves are shown. There is, however, an important 
difference with respect to the scenario depicted in panel (a). The initial part of the coexistence path is the same: 
the N$_{\rm r}$-N$_{\rm p}$ coexistence starts at $t=0$ and proceeds through the curves corresponding to 
increasing values of $t$. However, upon a further increase in $\eta_0$, the N$_{\rm p}$ phase undergoes a 
second-order transition to a B phase. 
The volume fraction of the B phase then
increases relative to that of the N$_{\rm r}$ phase, from $0.7$ (green square) to 
$1$ (red square), until the N$_{\rm r}$ phase eventually disappears 
from the sample at the end of the transition, and the B phase occupies the entire volume.  

Note that in the scenario shown in panel (b) the N$_{\rm r}$ is incipient both at the beginning and at end of the transition, while the N$_{\rm p}$ phase ultimately transforms into the B phase, which fills the entire sample. 
By contrast, 
in panel (a) the initially incipient phase N$_{\rm r}$ phase transforms into the B phase, which again occupies
the entire sample at the end of the transition.
Determining which of these two scenarios is actually realized
would require a detailed analysis of the free energies along the corresponding coexistence paths. We have not
performed such an analysis because these transition are metastable with respect to the NU phases. We therefore cannot rule out  
the existence of a metastable N$_{\rm r}$-N$_{\rm p}$-B three-phase coexistence in 
the region of the phase diagram where the N$_{\rm r}$-N$_{\rm p}$ and N$_{\rm r,p}$-B coexistence curves approach one another.
 
\section*{Acknowledgements}

E.V. and Y.M.-R. acknowledge financial support from Grants No. 
PID2023-148633NB-I00/AI and No. PID2021-126307NB-C21/MICIU/AEI/10.13039/501100011033/FEDER, UE, respectively. D.dlH. acknowledges support through 
the Heisenberg program of the Deutsche Forschungsgemeinschaft (DFG) under 
Project No. 550390029.

\end{document}